\documentstyle[preprint,epsf,aps]{revtex}

\newcommand{\be}{\begin{equation}}
\newcommand{\ee}{\end{equation}}
\newcommand{\ba}{\begin{array}}
\newcommand{\ea}{\end{array}}
\newcommand{\bea}{\begin{eqnarray}}
\newcommand{\eea}{\end{eqnarray}}
\newcommand{\bit}{\begin{itemize}}
\newcommand{\eit}{\end{itemize}}
\newcommand{\ben}{\begin{enumerate}}
\newcommand{\een}{\end{enumerate}}

\newcommand{\up}{\uparrow}
\newcommand{\down}{\downarrow}
\newcommand{\btm}{\bibitem}

\begin{document}

\title{\large \bf  Scaling analysis of a model Hamiltonian \\
for Ce$^{3+}$ impurity in a cubic metal}
\author{Tae-Suk Kim\cite{now} and D. L. Cox}
\address{Department of Physics, Ohio State University, Columbus, 
  Ohio 43210, USA}
\date{\today}
\maketitle
\begin{abstract}

 We introduce various exchange interactions 
in a model Hamiltonian for Ce$^{3+}$ ions in cubic symmetry 
with three configurations ($f^0$,$f^1$,$f^2$).
With the impurity pseudo spin $S_I=1/2$, our Hamiltonian includes: 
(i) One-channel $S_c=1/2$ Anderson model; 
(ii) Two-channel $S_c=1/2$ Anderson model; 
(iii) An unforseen one-channel $S_c=3/2$ Anderson model with a non-trivial
fixed point;
(iv) Mixing exchange interaction between the $\Gamma_{6,7}$ and 
the $\Gamma_8$ conduction electron partial wave states; 
(v) Multiple conduction electron partial wave states.
Using the third-order scaling (perturbative renormalization group) 
analysis, we study stability of various 
fixed points relevant to various exchange interactions for Ce$^{3+}$ ions 
in cubic symmetry. 
 
PACS Nos.  74.70.Vy, 74.65.+n, 74.70.Tx
\end{abstract}

\pacs{PACS numbers here}

\section{Introduction.}
 Since the introduction of the orbitally nondegenerate 
Anderson model\cite{anderson}, 
realistic extensions of this model have been considered\cite{hirst,nozbland}. 
Hirst\cite{hirst} described the Anderson model 
including the spin-orbit interaction and the crystal electric field effect 
based on the group theory. 
A detailed study of possible low temperature physics 
along this line was given by 
Nozi\`{e}res and Blandin\cite{nozbland} for the magnetic ions 
in normal metal hosts (primarily transition metal ions). 
Nozi\`{e}res and Blandin discovered the multichannel
Kondo models which have non-trivial fixed points in the overscreened  
cases, but concluded that the realization of this model 
has a difficulty in real materials due to the channel asymmetry. 
Cox introduced the two-channel quadrupolar Anderson model\cite{quad}
to explain the physics of UBe$_{13}$, and listed possible
crystal symmetries\cite{coxham} for the existence of 
the two-channel exchange interaction 
in rare earth and actinide compounds or alloys. 
However, a detailed study is required for the stability 
of the two-channel fixed point against possible symmetry-breaking
fields. 
Though in a different context, the metallic two-level system 
(electron-assisted atom tunneling in a double well)
was shown to map into the two-channel 
Kondo model\cite{tls}. 

In this paper, we study in detail a one-impurity
Anderson model for Ce$^{3+}$ with three configurations 
in cubic symmetry in the same spirit as
Nozi\`{e}res and Blandin did for the transition metal ions. Our study 
can be generalized to any $f$ atomic configurations in cubic symmetry.
As we shall show below, the inclusion of the detailed energy structure 
in the atomic orbital leads to rich physics. 
In summary, we find various exchange interactions with the impurity pseudo 
spin $S_I=1/2$:
(a) One-channel $S_c=1/2$ exchange interactions for $\Gamma_{6,7}$ conduction 
electrons (magnetic ``Kramers''' doublet orbitals); 
(b) Two-channel $S_c=1/2$ exchange interaction for $\Gamma_8$ conduction 
electrons (quartet of orbitals);
(c) One-channel $S_c=3/2$ exchange interaction for $\Gamma_8$ conduction 
electrons;
(d) Mixing exchange interactions between $\Gamma_{6,7}$ and $\Gamma_8$ 
conduction electrons.
In addition, (e) the multiple conduction electron partial wave states 
enhance the exchange couplings. 
In connection to the 
normal states of the cuprate superconductors\cite{cuprate} 
and the recent discoveries 
of non-Fermi liquid systems\cite{nfl} in heavy fermion materials, 
the relevant question is whether the ground states of the models (b) or 
(c) can be realized in the presence of other exchange interactions. 

We will answer this question using perturbative renormalization 
group (third-order scaling) arguments.   
We find three kinds of stable fixed points with the impurity spin 
$S_I=1/2$: one-channel Fermi liquid fixed points; 
two-channel non-Fermi liquid fixed point; and three-channel non-Fermi liquid
fixed points. In addition, a ``zoo" of unstable fixed points are discovered.
At these unstable fixed points, all the possible exchange interactions 
can exist in the octuplet manifolds of the conduction electrons. Furthermore 
the multiple conduction electron partial waves generate the multi-channel
unstable fixed points.   Some of our fixed points possess dimensionless 
couplings which are rigorously in the perturbative regime, while for 
the above mentioned one-,two-,and three-channel cases this is not so. 
Nonetheless, since we are only after the existence of the fixed points, 
the scaling theory has proven {\it qualitatively} accurate for non-trivial
two- and three-channel fixed points, and the one-channel fixed points 
are well known to flow to strong coupling.  Thus we have confidence in 
the correctness of our qualitative conclusions regarding stability of the
various fixed points, provided we understand the modification for the 
one-channel case. 

Similar results and conclusions 
have been reached by Zar\'{a}nd\cite{zarand} in the two-level system 
Kondo model (electron assisted tunneling of an atom in a double well)
where he employed a large number of channels (large real spin) and 
thus rigorously controlled the perturbative renormalization group theory. 
We have verified the essential correctness of our scaling results for
the $S_c=3/2$ conduction electrons with a numerical renormalization group 
study of the model (also combined with some conformal field theory analysis
of the model)\cite{tskim,klctobe}
A recent conformal field theory study of single channel, 
large conduction spin Kondo models, with special 
emphasis on the $S_c=3/2$ case; the results are compatible with ours
\cite{sengupta}.  Since this fixed point is ultimately unstable against
partial wave mixing exchange terms, it may be of little physical relevance.
However, an interesting theoretical question is raised by our study of 
the fixed point, bolstered by an independent numerical renormalization group
and conformal field theory study\cite{klctobe}.  Namely, the universality
class  of the fixed point is not that of the three-channel $S_I=1/2$ 
Kondo model as conjectured by Tsvelik and Wiegman\cite{tsvwieg} and
utilized in their Bethe-Ansatz treatment of the multichannel model, a 
point also noted by Zar\'{a}nd in his study of the novel unstable fixed
points of the TLS Kondo model\cite{zarand}.  
These ideas will be discussed more extensively in Ref. \cite{klctobe}.

Our paper is organized as follows:  Section II contains an overview of the
generalized Anderson 
model Hamiltonian and the underlying group theoretical motivations for 
understanding the various couplings which arise.  Section III contains 
a precise and systematic 
discussion of the hybridization matrix elements which feed into
the model together with derivation of the effective Schrieffer-Wolff couplings
between conduction and 4$f$ electrons upon integration out of virtual 
charge fluctuations.   Section IV contains the scaling analysis of 
various models which may be obtained in limiting cases of the full 
Hamiltonian including all possible conduction electron symmetry states
without adding in multiple partial waves.  Section V discusses the effect
of including multiple conduction electron partial waves in each symmetry 
channel.  Finally, we summarize and offer conclusions in Section VI.  
Three appendices detail finer points about the derivation of mixing 
matrix elements (App. A), the multiplicative renormalization group 
applied here (App. B), and the general form of the third order scaling
equations for our models (App. C).

\section{Model Hamiltonian I: Overview of Energetics and Symmetry.}

The model Hamiltonian we consider throughout this paper is the 
generalized Anderson model\cite{anderson,hirst}.  Schematically, this 
Hamiltonian has three parts, defined by 
\begin{equation}
H = H_{cond} + H_f + H_{hyb} 
\end{equation}
where $H_{cond}$ defines a broad flat conduction band of
width $D$, $H_f$ defines the quasi-atomic 4$f$ states on the
Ce site as modified by metallic screening of the direct Coulomb 
interaction and the crystalline electric field (CEF), 
and $H_{hyb}=H_{01}+H_{12}$
describes the hybridization between the 4$f$ and conduction electrons.
$H_{01}$ describes mixing of $f^0-f^1$ excitations with conduction 
electrons, and $H_{12}$ describes mixing of $f^1-f^2$ excitations
with conduction electrons.  
We shall describe $H_{hyb}$ in detail in the next section; in 
this section we shall focus on the symmetry and energetic aspects
of the first two terms.  In particular, we shall emphasize the 
symmetries of tensor operators from the conduction and $f$ sectors
which may be coupled together in the effective exchange interactions
mediated by virtual interconfiguration excitations.

Before proceeding further, we provide a brief physical description of the
different irrep labels of the cubic group:\\
(1) $\Gamma_6$ and $\Gamma_7$ are magnetic Kramers' doublet 
(Kramers'). Two degenerate states can be transformed into each other 
with time reversal symmetry operator on them and will be labeled with 
pseudo-spin $\up$ or $\down$. That is, the $\Gamma_6$ and $\Gamma_7$ 
CEF states are similiar to the $J=1/2$ angular momentum manifold.\\
(2) $\Gamma_8$ is a magnetic quartet ($\Gamma_8 = \Gamma_3\otimes \Gamma_7$). 
With the time reversal symmetry operator, the four-fold degeneracy can be 
decomposed into two pairs. These two pairs are disjoint in terms of 
time reversal and will be labeled by orbital indices, $\pm$, which
correspond to the quadrupolar or orbital 
$\Gamma_3$ degree of freedom -- stretched ($3z^2-r^2$ like)
or squashed ($x^2-y^2$ like) local orbitals guaranteed degenerate by
the cubic symmetry. The $\Gamma_7$ degree of 
freedom will be labeled as $\up$ and $\down$ as above. Overall, 
the $\Gamma_8$ quartet is labeled by $\pm \up/\down$.  Under certain 
circumstances it is more favorable to view the $\Gamma_8$ manifold as 
an effective $S=3/2$ manifold, as we shall discuss below.  \\
(3) The orbital singlets $\Gamma_1$ and $\Gamma_2$ do not need any further 
label.  The $\Gamma_2$ is a pseudoscalar, the lowest even parity version
being $(x^2-y^2)(y^2-z^2)(z^2-x^2)$. \\
(4) The orbital (non-magnetic, or non-Kramers') doublet $\Gamma_3$ will be
labeled by $\pm$ as in the above. In this case, time reversal symmetry 
operator does not transform one state to the other as for a Kramers'
doublet. \\
(5) The magnetic triplets $\Gamma_4$ and $\Gamma_5$ are labeled by $0, \pm 1$.
Under the time reversal symmetry operation, the $\pm 1$ states transform into 
the $\mp 1$ states, respectively. On the other hand, the state $0$ 
transforms into itself. That is, the $\Gamma_4$ and $\Gamma_5$ CEF
states are similiar to the $J=1$ multiplets under time reversal 
symmetry operation.

 In the absence of energetic considerations, there are 14 (91) 
different degenerate eigenstates in the $f^1$ ($f^2$) configuration. 
These states can be progressively split by the strong on-site 
Coulomb ``Hund's rule'' interactions ($f^2$), 
spin-orbit coupling, and the crystalline electric field 
(CEF). The energy hierarchy in Cerium is, roughly, that 
the interconfiguration splitting are several eV, 
the exchange splittings are about $0.2 - 0.5$ eV for the $f^2$,
the spin orbit splittings are order 0.3 eV, 
and the typical CEF splittings are from a few to tens of  meV.
Specifically, the $f^0-f^1$ interconfiguration excitation energy is 
$E(f^1)-E(f^0) \approx -2$eV, and the $f^1-f^2$ energy difference
is $E(f^2)-E(f^1) \approx 4$eV\cite{herbst}.  
Traditionally, $E(f^1)-E(f^0)$ is referred to as $\epsilon_f$ while
$E(f^2)-E(f^1)$ is referred to as $\epsilon_f + U$, meaning the 
effective Coulomb interaction is about 6 eV.  

 For the $f^1$ configuration, the energy level structure is simple 
due to the absence of the on-site Coulomb interaction. Spin-orbit coupling
first results in $J=5/2, 7/2$ multiplets. In most Ce$^{3+}$ compounds
and alloys, this energy splitting is always about 0.3 eV \cite{lssplit}.
Furthermore the crystalline electric field will
split each $J=5/2, 7/2$ multiplet. The CEF can mix two different
$J$'s but with the second order perturbation correction of order
$|<J=5/2 | H_{CEF} | J=7/2>|^2 / \Delta_{LS}$ which is of the order of 
$0.1 \times \Delta_{\rm CEF}$. 
Neglecting the mixing of two different $J$'s in the presence of the CEF, 
the irreducible representations  
(irreps) of the full rotation group, $D_{5/2}, D_{7/2}$ decompose in 
the cubic field as
\cite{group}
\bea
D_{5/2} &=& \Gamma_7 \oplus \Gamma_8, \\
D_{7/2} &=& \Gamma_6 \oplus \Gamma_7 \oplus \Gamma_8  ~.
\eea
Here $\Gamma_6$ and $\Gamma_7$ are the magnetic 
Kramers doublet's irreps and $\Gamma_8$ is the magnetic quartet irrep 
of the cubic double group. 
The eigenstates of $\Gamma_{7,8}$ for $J=5/2$ multiplets are given 
in Table I\cite{llw}.
We will remove the superscript $J=5/2$ below when the context is clear.
The eigenstates of $\Gamma_{6,7,8}$ for $J=7/2$ multiplets are listed 
in Table \ref{J3.5}.
The choice of the overall phase is arbitrary in defining the CEF eigenstates.
The above convention will make the effective exchange interaction 
a simple form.  Throughout this work we shall assume that the 
$J=5/2,\Gamma_7$ doublet lies lower than the $J=5/2,\Gamma_8$ quartet.  

 For the $f^2$ configuration, the existence of the strong on-site 
Coulomb interaction complicates the energy level structure.  
The LS Russel-Saunders scheme applies to the $f$ electron in 
Ce\cite{lsscheme}.
First in the absence of both spin-orbit coupling and CEF, total orbital
angular momentum $\vec{L}$ and total spin angular momentum $\vec{S}$
are good quantum numbers, and thus energy eigenstates can be sorted out
by these quantum numbers. 
In the presence of spin-orbit coupling, the total angular momentum
$\vec{J} = \vec{L} + \vec{S}$ is a good quantum number.
There are seven possible values of $J$ starting from 0 through 6. 
When Ce$^{3+}$ ions are sitting in a cubic environment, each $J$ 
multiplet further splits into CEF energy levels (see Table \ref{f2J}). 
For example, the $f^2$ Hund's rule ground multiplet $J=4 (^3H_4)$ decomposes 
as
\bea
 D_4 &=& \Gamma_1 \oplus \Gamma_3 \oplus \Gamma_4 \oplus \Gamma_5.
\eea
All in all, there are 7 $\Gamma_1$'s, 3 $\Gamma_2$'s, 9 $\Gamma_3$'s, 
9 $\Gamma_4$'s, and 12 $\Gamma_5$'s in the $f^2$ configuration.
Here $\Gamma_1$ and $\Gamma_2$ are singlet irreps, 
$\Gamma_3$ the non-Kramers doublet irrep (nonmagnetic), 
and $\Gamma_{4,5}$ the magnetic triplet irreps of the 
cubic point group. 
The eigenstates of $\Gamma_{1,3,4,5}$ for $J=4$ multiplets are in Table 
\ref{J4}

 We now consider the conduction electrons. According to the Anderson model
picture, the conduction electrons can hop on and off the atomic orbitals 
at the impurity site. 
The Bloch electrons may be projected at the impurity site into three irreps  
in the cubic symmetry: $\Gamma_6$, $\Gamma_7$, $\Gamma_8$.
For the $f$ shell in Ce, we expect the $l=3$ 
conduction electron partial waves are most strongly coupled to the $f$ shell.
For the isotropic hybridization, only the $l=3$ components can hybridize 
with the $f$ orbitals. In the presence of the spin-orbit coupling, the 
total angular momentum is a good quantum number. Thus we have 
$J=5/2, 7/2$ conduction electron partial waves. 
these $J$ multiplets further split into CEF irreps 
in crystal environments as mentioned above. 
When the real charge fluctuations are removed from the model system
in the Kondo limit, we have to construct the tensor operators representing
each CEF state for the $f^1$ configuration and the projected conduction 
electron CEF states. The relevant tensor operators are 
for $\Gamma_{6,7,8}$ CEF states. We list the decomposition of conduction 
electron operators into irreducible tensor opertors in Table \ref{tensor}
using the character table\cite{group}.
In the direct product, the first CEF states are written as ket, and the 
second as bra. The $\Gamma_{6,7}$ tensor operator ($2\times 2$ tensor) 
is a direct sum of a charge operator ($\Gamma_1$) and 
a pseudo spin operator ($\Gamma_4$). Indeed, 
the Schrieffer-Wolff transformation leads to two interaction 
terms: the spin exchange interaction and the pure potential scattering term. 
The relevant $\Gamma_8$ tensor operators are the $2\Gamma_4$ triplets 
for our model. In the conduction electron $\Gamma_8$ tensor space, 
one of the two
$\Gamma_4$'s gives rise to the pseudo spin $S_c=1/2$ operators with two 
degenerate orbital channels and the other to the pseudo spin $S_c=3/2$ 
operator with one channel.  The interpretation of the orbital channels
is that these are shape degrees of freedom of the local $\Gamma_8$
states--one is ``stretched'' or $3z^2-r^2$ like, and the other 
``squashed'' or $x^2-y^2$ like.  
Thus, the $\Gamma_{6,7}$ conduction electrons provide two different channels
with the pseudo spin $S_c=1/2$, while the $\Gamma_8$ conduction electrons 
provide two degenerate channels with $S_c=1/2$ or one channel with $S_c=3/2$.

 As a further complication, mixing is possible between $\Gamma_{6,7}$ and
$\Gamma_8$ conduction electrons. The mixing tensor operators can be read off
from Table \ref{tensor}. 
That is, each mixing can provide one pseudo spin representation. 
Hence from the group theory, we conclude that there are in total 6 possible 
exchange interactions between the $f^1J=5/2~\Gamma_7$ pseudo-spin and 
the three-symmetry conduction electrons. 
We are going to derive these in the next section.

\section{Model Hamiltonian II: Anderson hybridization.} 

In this section we derive explicitly the forms of the hybridization
matrix elements between excited $f^0,f^2$ states and $f^1$ states 
which are mediated by conduction electrons hopping on and off the 
impurity.  We also derive the contributions to the effective 
exchange interactions obtained by integrating out virtual charge
fluctuations to the $f^0,f^2$ states. 

 The conventional theory of Ce$^{3+}$ impurities assumes an infinite 
on-site Coulomb interaction, which removes the $f^2$ configuration from 
consideration. 
When we relax our assumption about the 
strong on-site Coulomb interaction and we include the detailed atomic energy
structure in cubic crystals, we can construct a variety of 
model Hamiltonians. In this work we assume that the magnetic doublet 
$f^1J=5/2~\Gamma_7$ states lies 
lower than the $f^1J=5/2~\Gamma_8$ quartet states. 
The one-channel $S_c=S_I=1/2$ Anderson hybridization 
is present between $f^0$ and 
$f^1J=5/2~\Gamma_7$ states.   This gives rise to the conventional 
single channel Kondo coupling.  
From the hybridization between $f^1$ and $f^2$ configurations, various exchange
interactions arise from virtual charge fluctuations, which include: \\
(i) the one-channel exchange interaction for magnetic doublet 
$\Gamma_{6,7}$ conduction 
electron orbitals;\\
(ii) the two-channel $S_c=1/2$ exchange interaction for $\Gamma_8$ conduction 
quartets;  \\
(iii) the one-channel $S_c=3/2$ exchange interactions for $\Gamma_8$ 
conduction electrons; \\
(iv) mixing exchange interactions 
between the $\Gamma_8$ and the $\Gamma_{6,7}$ conduction 
electrons; \\
(v) multiple conduction electron partial wave states.

 Below we list all the possible terms arising from the Anderson 
hybridization in the cubic symmetry. All the possible selection rules for 
the hybridization are listed in Table \ref{hyb}. Since the Anderson 
hybridization is irreducible tensor of $\Gamma_1$, the selection rule is 
solely determined by the three components: the conduction electron 
partial wave states and the other
two neighboring atomic configuration states. In the $f^1$ configuration, we 
only consider the ground CEF states $f^1J=5/2~\Gamma_7$. 
In the $f^2$ configuration, 
there are all $7 \Gamma_1$, $3 \Gamma_2$, $9 \Gamma_3$, $9 \Gamma_4$, and 
$12 \Gamma_5$ CEF states.

 For definiteness, we will consider isotropic hybridization.   By this
we mean that the original hybridization potential is taken as isotropic
and this will be projected down to the appropriate irreps of the cubic
point group symmetry at the Ce ion.  
Then $J_c=5/2, 7/2$ conduction electron partial 
waves are mediating the tunneling process between neighboring configurations
of the atomic orbitals. 

 The mixing between the $f^0$ and the $f^1$ configurations is
simple and straightforward to evaluate, 
since the $f^0$ is a singlet with $\Gamma_1$ 
character.   This is given by the Hamiltonian terms 
\bea
H_{01}^{(0)}
 &=& V ~ \sum_{\epsilon}\sum_{j_cm_c} ~ 
  c_{\epsilon j_cm_c}^{\dagger} ~|f^0><f^1;j_cm_c |
  + h.c. 
\eea
in the absence of the CEF, or 
\bea
H_{01}
 &=& V ~ \sum_{\epsilon}\sum_{j_c\Gamma_c d_c} ~ 
  c_{\epsilon j_c\Gamma_c d_c}^{\dagger} ~|f^0><f^1;j_c\Gamma_c d_c |
  + h.c. 
\eea
in the presence of the CEF.  
Here the conduction 
electron Bloch states are symmetry adapted into appropriate 
partial wave states labeled by cubic irrep indices $\Gamma_c$ 
at the impurity site. These projected
quantum labels play an important role in determining the ground state
properties of the model system. 

However, the mixing between the $f^1$ and the $f^2$ configurations becomes
complicated due to the fractional parentage and the Clebsch-Gordan 
coefficients. First we consider the case in the absence of the CEF.
\bea
H_{12}^{(0)} 
 &=& V ~\sum_{\epsilon} \sum_{j_cm_c}\sum_{j_1m_1} \sum_{LSJM} 
     \Lambda(j_cm_c;f^1j_1m_1| f^2LSJM) \nonumber\\
 && \times 
     c_{\epsilon j_cm_c}^{\dagger} ~|f^1j_1m_1><f^2LSJM| 
     + h.c. 
\eea
Here we use the reduced matrix elements are defined 
\bea
\Lambda(f^1j_1m_1; j_cm_c| f^2LSJM)
 &\equiv& <j_1m_1|~f_{j_cm_c}~|f^2LSJM> \nonumber\\
 &=& K (j_c; f^1j_1| f^2LSJ) ~ <j_cm_c; j_1m_1 | JM>. 
\eea
In the last line we used the Wigner-Eckart theorem. The prefactor
$K (j_c; f^1j_1| f^2LSJ)$ is the fractional parentage coefficient 
(see the Appendix \ref{mixing} for more details). 
Though we projected the atomic Fock space using the
symmetry quantum labels, these states are not energy eigenstates of the 
atomic Hamiltonian. The Coulomb (exchange) interaction will mix 
multiplets with the same $J$. This will introduce another unitary 
transformation from the 
symmetry states $|f^2LSJM>$ to the energy eigenstates. These energy 
eigenstates still respect the total angular momentum and are labeled by
$JM$ with extra energy quantum label. We eschew this complication in
favor of working in the simpler LS coupling limit, which is accurate
for Ce ions in any case.  

 As will be discussed below, the cubic Clebsch-Gordan coefficients for 
different $J$'s are equal to within a prefactor for $f^1~\Gamma_7$ CEF 
states. Hence the mixing between
the different multiplets with the same value of $J$ will not change
the matrix form (hybridization or mixing matrix).

 In passing we note that the same approach 
was undertaken for the $f^2$ and $f^3$ configurations including
only the Hund's rule ground multiplets\cite{japan}.  In principle, 
the discussion we give here for the $f^1,J=5/2,\Gamma_7$ doublet
should equally well apply to the $f^3,J=9/2,\Gamma_6$ doublet
considered in Ref. \cite{japan}, which however contends that only
a fermi liquid state is possible.   We shall defer a 
discussion about this different conclusion between the two works 
to the last section of the paper.  

We now turn to a systematic discussion of the hybridization matrix
elements between $f^1$ and $f^2$ and the 
effective exchange interactions mediated by virtual $f^2$ fluctuations.

\subsection{The hybridization between $f^1J=5/2~\Gamma_7$ and 
$f^2 \Gamma_1, \Gamma_2$.}
 As shown in Table \ref{hyb}, only the $\Gamma_7 (\Gamma_6)$ 
conduction electrons can mix with the $f^1\Gamma_7$ 
to reach the $f^2\Gamma_1 (\Gamma_2)$ states. 
The corresponding cubic Clebsch-Gordan 
coefficients can be written in a very simple form
\bea
<j_c\Gamma_7 \alpha; {5\over 2} \Gamma_7 \beta | J_2\Gamma_1>
 &=& (-1)^{1/2-\alpha} ~(\beta,\bar{\alpha})~ 
    <j_c\Gamma_7; {5\over 2} \Gamma_7| J_2\Gamma_1>, \\ 
<j_c\Gamma_6 \alpha; {5\over 2} \Gamma_7 \beta | J_2\Gamma_2>
 &=& (-1)^{1/2-\alpha} ~(\beta,\bar{\alpha})~ 
    <j_c\Gamma_7; {5\over 2} \Gamma_7| J_2\Gamma_2>. 
\eea
The phase dependence derives from time reversal symmetry. 
The common factors, which are independent of the CEF degeneracy labels,
$~~<j_c\Gamma_7; {5\over 2} \Gamma_7| J_2\Gamma_1>~~$ 
and $~~<j_c\Gamma_7; {5\over 2} \Gamma_7| J_2\Gamma_2>$ are listed in 
Table \ref{gamma1} and \ref{gamma2}.

Then the Anderson hybridization can be written as
\bea
H_{12} (\Gamma_1) 
 &=& V \sum_{\epsilon \alpha} \sum_{j_cLSJ} (-1)^{\alpha-1/2}
    ~K(j_c; f^1{5\over 2}~|~f^2 LS:J) ~
    < j_c \Gamma_7; {5\over 2} \Gamma_7 |J \Gamma_1>
    \nonumber\\
 && \times c_{\epsilon j_c \Gamma_7 \alpha}^{\dagger} 
    ~|f^1 {5\over 2}\Gamma_7 \bar{\alpha}><f^2LS:J \Gamma_1 | + h.c., \\
H_{12} (\Gamma_2) 
 &=& V \sum_{\epsilon \alpha} \sum_{j_cLSJ} (-1)^{\alpha-1/2}
    ~K(j_c; f^1{5\over 2}~|~f^2 LS:J) ~
    < j_c \Gamma_6; {5\over 2} \Gamma_7 |J \Gamma_2>
    \nonumber\\
 && \times c_{\epsilon j_c \Gamma_6 \alpha}^{\dagger} 
    ~|f^1 {5\over 2}\Gamma_7 \bar{\alpha}><f^2LS:J \Gamma_2 | + h.c.  
\eea
In the mixing, the degeneracy index for a time reveral pair is active, i.e.,
can change.  
In the local moment limit, the Schrieffer-Wolff transformation\cite{sw} 
leads to the {\it one-channel antiferromagnetic} exchange interaction 
\bea
H_1(\Gamma_1)
 &=& \sum_{ij=5/2,7/2} J_{ij} ~\vec{S}_{c\Gamma_7ij} (0) 
    \cdot \vec{S}_{\Gamma_7}, ~~J_{ij} = \sqrt{J_i J_j}, \\
J_{i=j_c} 
 &=& \sum_{LSJ} {2 |V|^2 \over E_2 (LSJ) - E_1} ~
    |K(j_c; f^1{5\over 2}~|~f^2 LS:J)|^2 ~
    |< j_c \Gamma_7; {5\over 2} \Gamma_7 |J \Gamma_1>|^2, \nonumber\\
 && 
\eea
and 
\bea
H_1 (\Gamma_2)
 &=& J ~\vec{S}_{c\Gamma_6} (0) \cdot \vec{S}_{\Gamma_7}, \\
J &=& \sum_{LSJ} {2 |V|^2 \over E_2 (LSJ) - E_1} ~
    |K({7\over 2}; f^1{5\over 2}~|~f^2 LS:J)|^2 ~
    |<{7\over 2} \Gamma_6; {5\over 2} \Gamma_7 |J \Gamma_2>|^2. \nonumber\\
 && 
\eea
Here the conduction pseudo-spin is $S_c = 1/2$. We also introduced
the energy levels for atomic states in the $f^1$ and the $f^2$ 
configurations. $E_1$ is the energy level for the $f^1~J=5/2~\Gamma_7$
state and $E_2 (LSJ)$ for the $f^2$ atomic state with quantum labels
of orbital ($L$), spin ($S$), and total angular momenta ($J$).  

\subsection{The hybridization between $f^1J=5/2~\Gamma_7$ and 
$f^2 \Gamma_3$.}
Only the $\Gamma_8$ conduction electrons can mix between 
the $f^1J=5/2 \Gamma_7$ and the $f^2 \Gamma_3$'s. 
The relevant cubic Clebsch-Gordan coefficients are 
\bea
<j_c \Gamma_8 n'\alpha'; {5\over 2} \Gamma_7 \alpha | J_2 \Gamma_3 n > 
 &=& (-1)^{\alpha+1/2} ~ (\alpha', \bar{\alpha}) ~(n',n)~
    <j_c \Gamma_8; {5\over 2} \Gamma_7 |J_2 \Gamma_3>. \nonumber\\
 && 
\eea    
The overall phase factor derives from time reversal symmetry.
The common factor $<j_c \Gamma_8; {5\over 2} \Gamma_7 |J_2 \Gamma_3>$
is listed in Table \ref{gamma3}.
Hence the hybridization can be written as
\bea
H_{12} (\Gamma_3) 
 &=& V \sum_{\epsilon j_cLSJ}\sum_{n=\pm} \sum_{\alpha=\up, \down} 
    (-1)^{\alpha-1/2}
    ~K(j_c; f^1{5\over 2}~|~f^2 LS:J) ~
    < j_c \Gamma_8; {5\over 2} \Gamma_7 |J \Gamma_3>
    \nonumber\\
 && \times c_{\epsilon j_c \Gamma_8 n \alpha}^{\dagger} 
    ~|f^1 {5\over 2}\Gamma_7 \bar{\alpha}><f^2LS:J \Gamma_3 n | + h.c.  
\eea
In the hybridization, the time reversal pair index $\alpha=\up, \down$ is 
active while the orbital index $n=\pm$ does not change and 
is not active. Thus the time reversal pair ($\up/\down$) gives rise to 
the pseudo-spin and the orbital doublet ($\pm$) provides two channels. 
The Schrieffer-Wolff transformation leads to the two-channel exchange
interaction
\bea
H_1
 &=& \sum_{ij=5/2,7/2} \sum_{n=\pm} J_{ij} ~\vec{S}_{c\Gamma_8ijn} (0) 
    \cdot \vec{S}_{\Gamma_7}, ~~J_{ij} = \sqrt{J_i J_j}, \\
J_{i=j_c} 
 &=& \sum_{LSJ} {2 |V|^2 \over E_2 (LSJ) - E_1} ~
    |K(j_c; f^1{5\over 2}~|~f^2 LS:J)|^2 ~
    |< j_c \Gamma_8; {5\over 2} \Gamma_7 |J \Gamma_3>|^2. \nonumber\\
 && 
\eea
Here the conduction pseudo-spin is $S_c = 1/2$.

\subsection{The hybridization between $f^1J=5/2~\Gamma_7$ and 
$f^2 \Gamma_4$.}
 The $\Gamma_7$ or $\Gamma_8$ conduction electrons can mix between the
$f^1J=5/2\Gamma_7$ and the $f^2 \Gamma_4$'s. The hybridization has the 
following form 
\bea
H_{12} (\Gamma_4) 
 &=& V \sum_{\epsilon} \sum_{j_cLSJ} 
    ~K(j_c; f^1{5\over 2}~|~f^2 LS:J) ~
    < j_c \Gamma_8; {5\over 2} \Gamma_7 |J \Gamma_4> \nonumber\\
 && \hspace{0.5cm} \times \left\{
  \ba{l} 
   - \sum_{\alpha} ~\sqrt{2} ~
     c_{\epsilon j_c\Gamma_8 -\alpha}^{\dagger} ~ |f^1 \bar{\alpha} >
     < f^2 LSJ\Gamma_4 0 |  \\
   + (~ c_{\epsilon j_c\Gamma_8 -\up}^{\dagger} ~ |f^1 \up > 
      - \sqrt{3}~ c_{\epsilon j_c\Gamma_8 +\down}^{\dagger} ~ 
      |f^1 \down > 
     ~) < f^2 LSJ\Gamma_4 1 | \\
   + (~ c_{\epsilon j_c\Gamma_8 -\down}^{\dagger} ~ |f^1 \down > 
      - \sqrt{3}~ c_{\epsilon j_c\Gamma_8 +\up}^{\dagger} ~ 
      |f^1 \up > 
     ~) < f^2 LSJ\Gamma_4 \bar{1} | 
  \ea
  \right\} 
  \nonumber\\
 && + V \sum_{\epsilon} \sum_{j_cLSJ} 
    ~K(j_c; f^1{5\over 2}~|~f^2 LS:J) ~
    < j_c \Gamma_7; {5\over 2} \Gamma_7 |J \Gamma_4> \nonumber\\
 && \hspace{0.5cm} \times \left\{ ~
  \ba{l}
   \sum_{\alpha}
     c_{\epsilon j_c\Gamma_7\alpha}^{\dagger} ~ |f^1 \bar{\alpha} >
     < f^2 LSJ\Gamma_4 0 |  \\
  + \sqrt{2}~c_{\epsilon j_c\Gamma_7\up}^{\dagger} ~ |f^1 \up > 
     < f^2 LSJ\Gamma_4 1 |  \\
  + \sqrt{2}~c_{\epsilon j_c\Gamma_7\down}^{\dagger} ~ |f^1 \down > 
       < f^2 LSJ\Gamma_4 \bar{1} | 
  \ea 
  \right\}. 
\eea
The hybridization looks more complicated compared to those for 
$\Gamma_1, \Gamma_2$ and $\Gamma_3$ CEF states. However, this
complication comes from our special labeling of the CEF degeneracy
as we shall show below. We list the numerical common factors 
$< j_c \Gamma_8; {5\over 2} \Gamma_7 |J_2 \Gamma_4>$ and
$<j_c\Gamma_8;{5\over 2}\Gamma_7|J_2\Gamma_4>$ in Table \ref{gamma4}.
Note that the explicit dependence on the degeneracy labels is left out
in extracting the prefactors. 

The Schrieffer-Wolff transformation leads to a new exchange interaction
model with the $S_c=3/2$ conduction electrons interacting with the impurity 
pseudo spin $S_I=1/2$ in the $\Gamma_8$ conduction electron space
\bea
H_1 &=& \sum_{ij} J_{ij} \vec{S}_{c\Gamma_8ij} (0) \cdot \vec{S}_{\Gamma_7}
  = {\bf J} \otimes \vec{S}_{c\Gamma_8} (0) \cdot \vec{S}_{\Gamma_7}, ~~
  J_{ij} = \sqrt{J_i J_j}, \\
J_{i=j_c} 
 &=& \sum_{LSJ} {2 |V|^2 \over E_2 (LSJ) - E_1} ~
    |K(j_c; f^1{5\over 2}~|~f^2 LS:J)|^2 ~
    |< j_c \Gamma_8; {5\over 2} \Gamma_7 |J \Gamma_4>|^2. \nonumber\\
 && 
\eea
Here $S_c=3/2$. The exchange coupling can be rewritten in the matrix form.
The corresponding representations of the angular momentum 
operators are ($|+\up>, |+\down>, |-\up>, |-\down>$)
\bea
S_c^{z} 
 &=& {1\over 2} ~\pmatrix{-3 & 0 & 0 & 0 \cr 0 & 3 & 0 & 0 \cr
                           0 & 0 & 1 & 0 \cr 0 & 0 & 0 & -1 }
  = \pmatrix{ -3 & 0 \cr 0 & 1} \otimes {1\over 2} ~ \sigma^{z}, \nonumber\\
S_c^{+} 
 &=& \pmatrix{0 & 0 & 0 & 0 \cr 0 & 0 & \sqrt{3} & 0 \cr
                           0 & 0 & 0 & -2 \cr \sqrt{3} & 0 & 0 & 0 }, ~~~
  S_c^{x} = \pmatrix{0 & \sqrt{3} \cr \sqrt{3} & -2} 
           \otimes {1\over 2} ~ \sigma^{x}, \nonumber\\
S_c^{-} 
 &=& \pmatrix{0 & 0 & 0 & \sqrt{3} \cr 0 & 0 & 0 & 0 \cr
                           0 & \sqrt{3} & 0 & 0 \cr 0 & 0 & -2 & 0 }, ~~~
  S_c^{y} = \pmatrix{0 & -\sqrt{3} \cr -\sqrt{3} & -2} 
           \otimes {1\over 2} ~ \sigma^{y}. 
\eea
We can show that these matrices transform into the canonical $SU(2)$ 
form after the rearrangement of the states: 
\bea
(|+\up>, |+\down>, |-\up>, |-\down>)
 &\rightarrow & (|{3\over 2}(-{3\over 2})>, |{3\over 2}{3\over 2}>,
      -|{3\over 2}{1\over 2}>, -|{3\over 2}(-{1\over 2})>). \nonumber
\eea
This extra complication comes from our biased labeling of the CEF degeneracy
toward the two-channel exchange interaction. 
>From a viewpoint of the two-channel exchange interaction, the induced 
exchange interaction can be written as
\bea
H_1 &=& {\bf J} \otimes (-\tau^{0} - 2 \tau^{i}) \otimes S_{\Gamma_8}^{i}
     ~ S_{\Gamma_7}^{i}, \\
\tau^{x} 
 &=& -{1\over 2}~\tau^{3} - {\sqrt{3} \over 2} ~ \tau^{1}, ~~
 \tau^{y} = -{1\over 2}~\tau^{3} + {\sqrt{3} \over 2} ~ \tau^{1}, ~~
 \tau^{z} = \tau^{3}. 
\eea
Thus we can see that the exchange interaction generated by the $f^2\Gamma_4$
CEF states are harmful to the two-channel Kondo effect. 

 On the other hand, the $\Gamma_7$ conduction electrons are coupled to the 
impurity spin with the ferromagnetic exchange coupling 
\bea
H_1 &=& -\sum_{ij} J_{ij} \vec{S}_{c\Gamma_7ij} (0) \cdot \vec{S}_{\Gamma_7},
  ~~ J_{ij} = \sqrt{J_i J_j}, \\
J_{i=j_c} 
 &=& \sum_{LSJ} {2 |V|^2 \over E_2 (LSJ) - E_1} ~
    |K(j_c; f^1{5\over 2}~|~f^2 LS:J)|^2 ~
    |< j_c \Gamma_7; {5\over 2} \Gamma_7 |J \Gamma_4>|^2. \nonumber\\
 && 
\eea
Here $S_c=1/2$. In this case, the exchange interaction, mediated by the 
$f^2\Gamma_4$ states, degrade the one-channel exchange coupling 
which is ever present between the $f^0$ and the $f^1$ configurations.

 Another complication arises through the $f^2\Gamma_4$ CEF states. Two
conduction electrons of $\Gamma_7$ and $\Gamma_8$ can mix with each other.
Since it can be easily generalized to the multiple conduction partial waves, 
we present the model for the one partial conduction wave case. 
In the $\Gamma_7\oplus \Gamma_8$ conduction electron space, 
all the possible exchange interactions are
\bea
H_1 
 &=& J^{i} \otimes S_c^{i} ~ S_{\Gamma_7}^{i}, \\
J^{i} 
 &=& J_7 \kappa^{0} + J_8 \tau^{0} + J_{78} \kappa^{i} 
   + J_{88} \tau^{i}, \nonumber\\ 
\kappa^0
 &=& \pmatrix{1 & 0 & 0 \cr 0 & 0 & 0 \cr 0 & 0 & 0}, ~~
\kappa^{x} = -{1\over 2} ~ \pmatrix{0 & \sqrt{3} & 1 \cr \sqrt{3} & 0 & 0 \cr
                                    1 & 0 & 0}, \nonumber\\
\kappa^{y} 
 &=& -{1\over 2} ~ \pmatrix{0 & -\sqrt{3} & 1 \cr -\sqrt{3} & 0 & 0 \cr
                                    1 & 0 & 0}, ~~
\kappa^{z}
 = \pmatrix{0 & 0 & 1 \cr 0 & 0 & 0 \cr 1 & 0 & 0}, \\    
\tau^0
 &=& \pmatrix{0 & 0 & 0 \cr 0 & 1 & 0 \cr 0 & 0 & 1}, ~~
\tau^{x} = -{1\over 2} ~ \pmatrix{0 & 0 & 0 \cr 0 & 1 & \sqrt{3} \cr
                                    0 & \sqrt{3} & -1}, \nonumber\\
\tau^{y} 
 &=& -{1\over 2} ~ 
   \pmatrix{0 & 0 & 0 \cr 0 & 1 & -\sqrt{3} \cr 0 & -\sqrt{3} & -1 }, ~~
\tau^{z}
 = \pmatrix{0 & 0 & 0 \cr 0 & 1 & 0 \cr 0 & 0 & -1}.     
\eea

\subsection{The hybridization between $f^1J=5/2~\Gamma_7$ and 
$f^2 \Gamma_5$.}
 The $\Gamma_6$ or $\Gamma_8$ conduction electrons can mix between the
$f^1J=5/2\Gamma_7$ and the $f^2 \Gamma_5$'s. The hybridization has the 
following form
\bea
H_{12} (\Gamma_5) 
 &=& V \sum_{\epsilon} \sum_{j_cLSJ} 
    ~K(j_c; f^1{5\over 2}~|~f^2 LS:J) ~
    < j_c \Gamma_8; {5\over 2} \Gamma_7 |J \Gamma_5> \nonumber\\
 && \hspace{0.5cm} \times ~\left\{ 
  \ba{l} 
   \sum_{\alpha} ~ \sqrt{2}~
     c_{\epsilon j_c\Gamma_8+\alpha}^{\dagger} ~ |f^1 \bar{\alpha} >
     < f^2 LSJ\Gamma_5 0 | \\
  + (~\sqrt{3}~c_{\epsilon j_c\Gamma_8-\up}^{\dagger}~|f^1 \up > 
     + c_{\epsilon j_c\Gamma_8 +\down}^{\dagger} ~ |f^1 \down > 
    ~) < f^2 LSJ\Gamma_5 1 | \\
  + (~\sqrt{3} ~ c_{\epsilon j_c\Gamma_8 -\down}^{\dagger} ~|f^1 \down > 
     + c_{\epsilon j_c\Gamma_8 +\up}^{\dagger} ~ |f^1 \up > 
    ~) < f^2 LSJ\Gamma_5 \bar{1} |  
 \ea
 \right\} 
 \nonumber\\
 && + V \sum_{\epsilon} \sum_{j_cLSJ} 
    ~K(j_c; f^1{5\over 2}~|~f^2 LS:J) ~
    < j_c \Gamma_6; {5\over 2} \Gamma_7 |J \Gamma_5> \nonumber\\
 && \hspace{0.5cm} \times \left\{ 
  \ba{l}
  - \sum_{\alpha} 
      c_{\epsilon j_c\Gamma_6\alpha}^{\dagger} ~ |f^1 \bar{\alpha} >
      < f^2 LSJ\Gamma_5 0 | \\
  + \sqrt{2} ~c_{\epsilon j_c\Gamma_6\up}^{\dagger} ~ |f^1 \up > 
      < f^2 LSJ\Gamma_5 \bar{1} | \\
  + \sqrt{2} ~c_{\epsilon j_c\Gamma_6\down}^{\dagger} ~ |f^1 \down > 
       < f^2 LSJ\Gamma_5 1 |  
  \ea
  \right\}.
\eea
The Schrieffer-Wolff transformation leads to a new exchange interaction
with the $S_c=3/2$ conduction electrons interacting with the impurity 
pseudo spin $S_I=1/2$ in the $\Gamma_8$ conduction electron space
\bea
H_1 &=& \sum_{ij} J_{ij} \vec{S}_{c\Gamma_8ij} (0) \cdot 
   \vec{S}_{\Gamma_7}, ~~J_{ij} = \sqrt{J_i J_j}, \\
J_{i=j_c} 
 &=& \sum_{LSJ} {2 |V|^2 \over E_2 (LSJ) - E_1} ~
    |K(j_c; f^1{5\over 2}~|~f^2 LS:J)|^2 ~
    |< j_c \Gamma_8; {5\over 2} \Gamma_7 |J \Gamma_4>|^2. \nonumber\\
 &&
\eea
Here $S_c=3/2$. This conduction electron pseudo spin is different from that 
defined in the previous section. 
The corresponding representations of the angular momentum 
operators are ($|+\up>, |+\down>, |-\up>, |-\down>$)
\bea
S_c^{z} 
 &=& {1\over 2} ~\pmatrix{1 & 0 & 0 & 0 \cr 0 & -1 & 0 & 0 \cr
                           0 & 0 & -3 & 0 \cr 0 & 0 & 0 & 3 }
  = \pmatrix{ 1 & 0 \cr 0 & -3} \otimes {1\over 2} ~ \sigma^{z}, \nonumber\\
S_c^{+} 
 &=& \pmatrix{0 & -2 & 0 & 0 \cr 0 & 0 & -\sqrt{3} & 0 \cr
                           0 & 0 & 0 & 0 \cr -\sqrt{3} & 0 & 0 & 0 }, ~~~
  S_c^{x} = \pmatrix{-2 & -\sqrt{3} \cr -\sqrt{3} & 0} 
           \otimes {1\over 2} ~ \sigma^{x}, \nonumber\\
S_c^{-} 
 &=& \pmatrix{0 & 0 & 0 & -\sqrt{3} \cr -2 & 0 & 0 & 0 \cr
                           0 & -\sqrt{3} & 0 & 0 \cr 0 & 0 & 0 & 0 }, ~~~
  S_c^{y} = \pmatrix{-2 & \sqrt{3} \cr \sqrt{3} & 0} 
           \otimes {1\over 2} ~ \sigma^{y}. 
\eea
We can show that these matrices transform into the canonical form after
the rearrangement of the states: 
\bea
(|+\up>, |+\down>, |-\up>, |-\down>)
 &\rightarrow & (-|{3\over 2}{1\over 2}>, |{3\over 2}(-{1\over 2})>,
      -|{3\over 2}(-{3\over 2})>, |{3\over 2}{3\over 2}>). \nonumber
\eea
>From a viewpoint of the two-channel exchange, the induced exchange  
interaction can be written as
\bea
H_1 &=& {\bf J} \otimes (-\tau^{0} + 2 \tau^{i}) \otimes S_{\Gamma_8}^{i}
     ~ S_{\Gamma_7}^{i}, \\
\tau^{x} 
 &=& -{1\over 2}~\tau^{3} - {\sqrt{3} \over 2} ~ \tau^{1}, ~~
 \tau^{y} = -{1\over 2}~\tau^{3} + {\sqrt{3} \over 2} ~ \tau^{1}, ~~
 \tau^{z} = \tau^{3}.
\eea
Thus in the $\Gamma_8$ conduction electron 
space, there are three competitors: two-channel Kondo effect and the other 
two $S_c=3/2$ Kondo effect (overscreened). 

 On the other hand, the $\Gamma_6$ conduction electrons are coupled to the 
impurity pseudo spin with the ferromagnetic exchange coupling 
\bea
H_1 
 &=& -\sum_{ij} J_{ij} \vec{S}_{c\Gamma_6ij} (0) \cdot \vec{S}_{\Gamma_7},~~
  J_{ij} = \sqrt{J_i J_j}, \\
J_{i=j_c} 
 &=& \sum_{LSJ} {2 |V|^2 \over E_2 (LSJ) - E_1} ~
    |K(j_c; f^1{5\over 2}~|~f^2 LS:J)|^2 ~
    |< j_c \Gamma_6; {5\over 2} \Gamma_7 |J \Gamma_4>|^2. \nonumber\\
 && 
\eea
Here $S_c=1/2$. In this case, the exchange interaction, mediated by the 
$f^2\Gamma_4$ states, degrade the one-channel exchange coupling. 

 Another complication arises through the $f^2\Gamma_5$ CEF states. Two
conduction electrons of $\Gamma_6$ and $\Gamma_8$ can mix with each other.
In the $\Gamma_6\otimes \Gamma_8$ conduction electron space, 
all the possible exchange interactions are
\bea
H_1 
 &=& J^{i} \otimes S_c^{i} ~ S_{\Gamma_7}^{i}, \\
J^{i} 
 &=& J_6 \gamma^{0} + J_8 \tau^{0} + J_{68} \gamma^{i} 
   + J_{88} \tau^{i}, \\ 
\gamma^0
 &=& \pmatrix{1 & 0 & 0 \cr 0 & 0 & 0 \cr 0 & 0 & 0}, ~~
\gamma^{x} = -{1\over 2} ~ \pmatrix{0 & 1 & -\sqrt{3} \cr 1 & 0 & 0 \cr
                                    -\sqrt{3} & 0 & 0}, \nonumber\\
\gamma^{y} 
 &=& -{1\over 2} ~ \pmatrix{0 & 1 & \sqrt{3} & \cr 1 & 0 & 0 \cr
                                    \sqrt{3} & 0 & 0}, ~~
\gamma^{z}
 = \pmatrix{0 & 1 & 0 \cr 1 & 0 & 0 \cr 0 & 0 & 0}, \\    
\tau^0
 &=& \pmatrix{0 & 0 & 0 \cr 0 & 1 & 0 \cr 0 & 0 & 1}, ~~
\tau^{x} = -{1\over 2} ~ \pmatrix{0 & 0 & 0 \cr 0 & 1 & \sqrt{3} \cr
                                    0 & \sqrt{3} & -1}, \nonumber\\
\tau^{y} 
 &=& -{1\over 2} ~ 
   \pmatrix{0 & 0 & 0 \cr 0 & 1 & -\sqrt{3} \cr 0 & -\sqrt{3} & -1 }, ~~
\tau^{z}
 = \pmatrix{0 & 0 & 0 \cr 0 & 1 & 0 \cr 0 & 0 & -1}.     
\eea

\subsection{The $\Gamma_6 \oplus \Gamma_7 \oplus \Gamma_8$ 
conduction electron space.}  
 In cubic symmetry, when we keep the lowest lying $f^1J=5/2~\Gamma_7$ 
CEF states, the most general exchange interaction form to each conduction 
electron partial wave species is 
\bea
H_1 &=& J^i \otimes S_c^{i} ~ S_{\Gamma_7}^{i}, \\
J^{i}
 &=& J_6 ~\gamma^{0} + J_7 \kappa^{0} + J_8 \tau^{0} 
     + J_{68}~\gamma^{i} + J_{78}~\kappa^{i} + J_{88} ~ \tau^{i}, \\
\gamma^0
 &=& \pmatrix{1 & 0 & 0 & 0 \cr 0 & 0 & 0 & 0 \cr 0 & 0 & 0 & 0 \cr
      0 & 0 & 0 & 0 }, ~~
\gamma^{x} = -{1\over 2} ~ 
  \pmatrix{0 & 0 & 1 & -\sqrt{3} \cr 0 & 0 & 0 & 0 \cr 
           1 & 0 & 0 & 0 \cr -\sqrt{3} & 0 & 0 & 0}, \nonumber\\
\gamma^{y} 
 &=& -{1\over 2} ~ 
  \pmatrix{0 & 0 & 1 & \sqrt{3} & \cr 0 & 0 & 0 & 0 \cr 
           1 & 0 & 0 & 0 \cr \sqrt{3} & 0 & 0 & 0}, ~~
\gamma^{z}
 = \pmatrix{0 & 0 & 1 & 0 \cr 0 & 0 & 0 & 0 \cr 
            1 & 0 & 0 & 0 \cr 0 & 0 & 0 & 0}, \\    
\kappa^0
 &=& \pmatrix{0 & 0 & 0 & 0 \cr 0 & 1 & 0 & 0 \cr 
              0 & 0 & 0 & 0 \cr 0 & 0 & 0 & 0}, ~~
\kappa^{x} = -{1\over 2} ~ 
  \pmatrix{0 & 0 & 0 & 0 \cr 0 & 0 & \sqrt{3} & 1 \cr 
           0 & \sqrt{3} & 0 & 0 \cr 0 & 1 & 0 & 0}, \nonumber\\
\kappa^{y} 
 &=& -{1\over 2} ~ 
  \pmatrix{0 & 0 & 0 & 0 \cr 0 & 0 & -\sqrt{3} & 1 \cr 
           0 & -\sqrt{3} & 0 & 0 \cr 0 & 1 & 0 & 0}, ~~
\kappa^{z}
 = \pmatrix{0 & 0 & 0 & 0 \cr 0 & 0 & 0 & 1 \cr 
            0 & 0 & 0 & 0 \cr 0 & 1 & 0 & 0}, \\    
\tau^0
 &=& \pmatrix{0 & 0 & 0 & 0 \cr 0 & 0 & 0 & 0 \cr 
              0 & 0 & 1 & 0 \cr 0 & 0 & 0 & 1}, ~~
\tau^{x} = -{1\over 2} ~ 
 \pmatrix{0 & 0 & 0 & 0 \cr 0 & 0 & 0 & 0 \cr 
          0 & 0 & 1 & \sqrt{3} \cr 0 & 0 & \sqrt{3} & -1}, \nonumber\\
\tau^{y} 
 &=& -{1\over 2} ~ 
   \pmatrix{0 & 0 & 0 & 0 \cr 0 & 0 & 0 & 0 \cr 
            0 & 0 & 1 & -\sqrt{3} \cr 0 & 0 & -\sqrt{3} & -1 }, ~~
\tau^{z}
 = \pmatrix{0 & 0 & 0 & 0 \cr 0 & 0 & 0 & 0 \cr 
            0 & 0 & 1 & 0 \cr 0 & 0 & 0 & -1}.     
\eea
The matrix algebra for $\gamma, \kappa, \tau$ is listed below.
We note that in the subspaces, $\Gamma_7\oplus\Gamma_8$ or 
$\Gamma_6\oplus\Gamma_8$, the exchange interactions are self-contained. No 
new interactions are generated with renormalization. 
Also in the whole $\Gamma_6 \oplus \Gamma_7 \oplus \Gamma_8$ conduction
electron space, the exchange interactions are self-contained. 

\bea
\gamma^{i} \gamma^{i} 
 &=& \gamma^{0} + {1\over 2} ~ \tau^{0} + {1\over 2} ~ \tau^{i}, \\
\gamma^{i} + \gamma^{j} 
 &=& - \gamma^{k} ~~(\mbox{cyclic}), \\
\gamma^{0} \gamma^{i} + \gamma^{i} \gamma^{0} 
 &=& \gamma^{i}, \\
\gamma^{i} \gamma^{j} + \gamma^{j} \gamma^{i} 
 &=& - \gamma^0 - {1\over 2}~\tau^{0} + \tau^{k} ~~(\mbox{cyclic}).
\eea
\bea
\gamma^{0}\tau^{i} + \tau^{i}\gamma^{0} 
 &=& 0, ~~\mbox{for}~~i=0,x,y,z, \\
\tau^{0}\gamma^{i} + \gamma^{i}\tau^{0}
 &=& \gamma^{i}, \\
\tau^{i}\gamma^{j} + \gamma^{j}\tau^{i} 
 &=& \gamma^{k} ~~(\mbox{cyclic}), \\
\gamma^{i}\tau^{j} + \tau^{j}\gamma^{i}
 &=& \gamma^{k} ~~(\mbox{cyclic}), \\
\gamma^{i}\tau^{i} + \tau^{i}\gamma^{i}
 &=& \gamma^{i}. 
\eea
\bea                                                         
&& \tau^{i} \tau^{i} = \tau^{0}, \\
&& \tau^{0} \tau^{i} + \tau^{i} \tau^{0} = 2 \tau^{i}, \\
&& \tau^{i} + \tau^{j} = - \tau^{k} ~~(\mbox{cyclic}), \\
&& \tau^{x}\tau^{y} = \tau^{y}\tau^{z} = \tau^{z} \tau^{x} 
  = - {1\over 2} ~ \tau^{0} - {\sqrt{3} \over 2} ~ \tau^{2}, \\
&& \tau^{y}\tau^{x} = \tau^{z}\tau^{y} = \tau^{x} \tau^{z} 
  = - {1\over 2} ~ \tau^{0} + {\sqrt{3} \over 2} ~ \tau^{2}. 
\eea
\bea
\kappa^{i} \kappa^{i} 
 &=& \kappa^{0} + {1\over 2} ~ \tau^{0} - {1\over 2} ~ \tau^{i}, \\
\kappa^{i} + \kappa^{j} 
 &=& - \kappa^{k} ~~(\mbox{cyclic}), \\
\kappa^{0} \kappa^{i} + \kappa^{i} \kappa^{0} 
 &=& \kappa^{i}, \\
\kappa^{i} \kappa^{j} + \kappa^{j} \kappa^{i} 
 &=& - \kappa^0 - {1\over 2}~\tau^{0} - \tau^{k} ~~(\mbox{cyclic}).
\eea
\bea
\kappa^{0}\tau^{i} + \tau^{i}\kappa^{0} 
 &=& 0, ~~\mbox{for}~~i=0,x,y,z, \\
\tau^{0}\kappa^{i} + \kappa^{i}\tau^{0}
 &=& \kappa^{i}, \\
\tau^{i}\kappa^{j} + \kappa^{j}\tau^{i} 
 &=& - \kappa^{k} ~~(\mbox{cyclic}), \\
\kappa^{i}\tau^{j} + \tau^{j}\kappa^{i}
 &=& - \kappa^{k} ~~(\mbox{cyclic}), \\
\kappa^{i}\tau^{i} + \tau^{i}\kappa^{i}
 &=& - \kappa^{i}. 
\eea
\bea
\kappa^{0}\gamma^{i} + \gamma^{i}\kappa^{0} 
 &=& 0, \\
\gamma^{0}\kappa^{i} + \kappa^{i}\gamma^{0} 
 &=& 0, \\
\kappa^{i}\gamma^{j} + \gamma^{j}\kappa^{i} 
 &=& {\sqrt{3} \over 2} ~ \chi, \\
\gamma^{i}\kappa^{j} + \kappa^{j}\gamma^{i} 
 &=& - {\sqrt{3} \over 2} ~ \chi, \\
\kappa^i \gamma^i + \gamma^i \kappa^i 
 &=& 0, \\
\kappa^{0}\chi + \chi\kappa^{0} 
 &=& \chi, \\
\gamma^{0}\chi + \chi\gamma^{0} 
 &=& \chi. 
\eea
Here $\chi$ is given by 
\bea
\chi &=& \pmatrix{0&1&0&0\cr 1&0&0&0 \cr 0&0&0&0 \cr 0&0&0&0}.
\eea
Though this new term ($\chi$) arises in the intermediate calculation step, 
this term cancels out in the final result.

 In deriving the scaling equations, the following relations are
useful 
\bea
J^{i}J^{j}+J^{j}J^{i}
 &=& (2J_6^2-J_{68}^2)~\gamma^0 + (2J_7^2-J_{78}^2)~\kappa^0 
    + \left[~2J_8^2-J_{88}^2-{1\over 2} ~(J_{68}^2+J_{78}^2)~\right]~\tau^0 
    \nonumber\\
 && + J_{68}~(~2J_{88} - J_6 - J_8~)~\gamma^{k} 
    + J_{78}~(~-2J_{88} - J_7 - J_8~)~\kappa^{k} \nonumber\\
 && + (~J_{68}^2 - J_{78}^2 - 2 J_8J_{88}~)~\tau^{k}.
\eea
Here $(i,j,k)$ is cyclic,  and
\bea
J^{i}J^{i}
 &=& (J_6^2+J_{68}^2)~\gamma^0 + (J_7^2+J_{78}^2)~\kappa^0 
    + \left[~J_8^2+J_{88}^2+{1\over 2} ~(J_{68}^2+J_{78}^2)~\right]~\tau^0 
    \nonumber\\
 && + J_{68}~(~J_{88}+J_6+J_8~)~\gamma^{i} 
    + J_{78}~(~-J_{88}+J_7+J_8~)~\kappa^{i} \nonumber\\
 && + \left[~2J_8J_{88} +{1\over 2}~(J_{68}^2-J_{78}^2~)~\right]~ 
       \tau^{i}.
\eea

\subsection{Summary}
 In summary, we discovered various exchange interactions.
In particular, a new type of exchange interaction is discovered: the $S_c=3/2$
conduction electrons interact with the impurity pseudo spin $S_I=1/2$.
All the possible exchange interactions in our model study are 
summarized in Fig.~\ref{modelfig}.

 The $\Gamma_6$ conduction electrons (pseudo spin $S_c=1/2$) 
are coupled to the impurity pseudo spin
$S_I=1/2$ with the {\it antiferromagnetic} exchange coupling through 
$f^2\Gamma_2$ states, and with the {\it ferromagnetic} exchange coupling 
through $f^2\Gamma_5$ states. The mixing between $\Gamma_6$ and $\Gamma_8$ 
conduction electrons is also present. That is, the $\Gamma_6$ conduction 
electrons can change their symmetry states ($\Gamma_8$) after scattered 
off the impurity pseudo spin.

 The $\Gamma_7$ conduction electrons (pseudo spin $S_c=1/2$)
are coupled to the impurity pseudo spin
$S_I=1/2$ with the {\it antiferromagnetic} exchange coupling through 
$f^2\Gamma_1$ states, and with the {\it ferromagnetic} exchange coupling 
through $f^2\Gamma_4$ states. 
The mixing between $\Gamma_7$ and $\Gamma_8$ conduction electrons 
is also present. That is, the $\Gamma_7$ conduction electrons can change 
their symmetry states ($\Gamma_8$) after scattered off the impurity spin.

 The $\Gamma_8$ conduction electrons can generate two different conduction
electron pseudo-spins: $S_c=1/2$ with two degenerate orbital channels and and 
$S_c=3/2$ with one channel. The $\Gamma_8$ conduction electrons are coupled
to the impurity pseudo spin in a two-channel exchange interaction through 
$f^2\Gamma_3$, and in a one-channel exchange interaction through 
$f^2\Gamma_{4,5}$ states. Both models lead to the overcompensation of the 
impurity pseudo spin $S_I=1/2$. The two-channel case has been studied 
extensively using several techniques and 
leads to a non-Fermi liquid fixed point.
Using the numerical renormalization group and the conformal field theory,
the one-channel $S_c=3/2$ case is also shown to lead to 
the non-Fermi liquid fixed point\cite{klctobe}.

 We can give some plausible argument why some of exchange couplings
are ferromagnetic while the others are antiferromagnetic. 
This argument follows from the addition of two angular momenta.
As noted before, the CEF manifolds in the $f^2$ configuration
can be roughly evisaged as angular momentum multiplets.

To hop into the orbital singlet $\Gamma_1$
($\Gamma_2$) in the $f^2$ configuration, 
the $\Gamma_7$($\Gamma_6$) conduction electrons 
have to be in the opposite 
pseudo-spin state compared to the $f^1 \Gamma_7$ (impurity) pseudo 
spin state. Thus the coupling should be {\it antiferromagnetic}.

Now we consider the orbital doublet $\Gamma_3$ in the $f^2$ 
configuration. This orbital doublet is also represented as
$J=0$ singlet in terms of time reversal. In this case,
the relevant $\Gamma_8$ conduction electrons are represented 
by pseudo-spin $S_c=1/2$ and should be in the opposite pseudo 
spin state compared to the $f^1 \Gamma_7$ (impurity) pseudo 
spin state. Thus the coupling should be {\it antiferromagnetic}.
>From this argument, we can clearly see why the $\Gamma_8$
conduction electron cannot be represented by pseudo-spin 
$S_c=3/2$ for the virtual charge fluctuation to the 
$\Gamma_3$ CEF states. 

When we consider the magetic triplets $\Gamma_4$ and $\Gamma_5$ 
in the $f^2$ configuration, the exchange coupling can be 
ferromagnetic or antiferromagnetic depending on 
which CEF symmetry electrons hop into the atomic orbitals.
These magnetic triplets can be represented by the $J=1$
manifolds. 
For the $\Gamma_6$ or the $\Gamma_7$ conduction electrons
which are represented by pseudo-spin $S_c=1/2$, the 
coupling should be {\it ferromagnetic} to form the 
$J=1$ triplet state from the addition of $S_c=1/2$
and $S_I=1/2$. As we noted before, the $\Gamma_8$ conduction 
electrons can be represented in both ways:
$S_c=3/2$ with one orbital channel or $S_c=1/2$ with 
two orbitally degenerate channels. Hence depending on
the spin representation, the coupling has different sign.
With $S_c=1/2$ representation, the coupling should be 
{\it ferromagnetic} as explained above. This picture in fact 
agrees with our results that the virtual charge fluctuations
to the triplets $\Gamma_4$ and $\Gamma_5$ in the $f^2$ are
harmful to the two-channel Kondo effect for the $\Gamma_8$
conduction electrons.  
With $S_c=3/2$ representation, the coupling should be
{\it antiferromagnetic} which can be deduced from the 
angular momentum addition of $S_c=3/2$ and $S_I=1/2$
which has to be $J=1$.

\section{Scaling Analysis I: Various fixed points}
 In this section we analyze our various model exchange interaction terms
using the third order scaling argument - perturbative renormalization
group (RG)\cite{nozbland,scale1,scale2,scale3}.
At temperature $T$, only the conduction electrons (thermally excited)
inside the band of order $T$ with respect to the Fermi level play an
important role in determining physical properties.
Thus we can integrate out the band edge states (virtually excited states)
to find the effective Hamiltonian. Though the following analysis is 
restricted to the perturbative regime (weak coupling limit),
we can derive qualitatively correct
results out of this. For quantitative results,
a full numerical renormalization group (NRG) study is required.

Our strategy is as follows: 
we start with each exchange interaction term and identify the relevant 
fixed points. After that, we include other interaction terms step by step
to see if the relevant fixed points remain stable. 
Finally, in the next section, 
 we consider the effect of the multiple conduction electron 
partial wave states on the various Kondo effects.

 In summary, we find that three types of fixed points are stable in 
the presence of the full exchange interactions: \\
(i) one-channel Fermi liquid fixed points ($\Gamma_6$ 
and $\Gamma_7$ conduction electrons); \\
(ii) two-channel non-Fermi liquid fixed point ($\Gamma_8$ conduction
electrons);\\
(iii) three-channel non-Fermi liquid fixed points (some linear 
combination of $\Gamma_8$ and $\Gamma_6$ or $\Gamma_7$
conduction electrons).  
In addition to these stable fixed points, we find a ``zoo" of unstable
fixed points at which various exchange interactions we can imagine 
are realized. 
When we consider the multiple conduction electron partial
waves, the relevant stable fixed points remain stable with enhanced
bare exchange couplings and new unstable multichannel fixed points 
are generated. 
 
\subsection{Structure of the fixed points}
 Before discussing the scaling equations for various exchange interactions,
we wish to establish some mathematical structure of the fixed points
generated by the third order scaling equations. We consider the following
generic exchange interactions
\bea
H_1 &=& J^{i} \otimes S_c^{i} (0) ~ S_I^{i}.
\eea
Here $S_c=S_I=1/2$ are spin operators and $J^{i}$ is the exchange coupling 
matrices whose dimension is $4\times 4$ in our model case.
We first consider some arbitrary $SU(2)$ angular momentum operators
\bea
L^{i} &\equiv & M^{i} \otimes {1\over 2} ~\sigma^{i}.
\eea
Here $\sigma^{i}$ are Pauli matrices and $M^{i}$ are undetermined at this
point. Now we demand that $L^{i}$ satisfy the $SU(2)$ spin algebra, that is,
\bea
~[~L^{i}, ~L^{j}~]
 &=& i\epsilon_{ijk} ~ L^{k}.
\eea 
Here $\epsilon_{ijk}$ is Levi-Civita antisymmetry tensor. Then we find the 
condition for $M^{i}$'s to satisfy
\bea
\{ ~M^{i}, ~M^{j}~\} &=& 2~M^{k}.
\eea
Here $(i,j,k)$ is cyclic. Up to this point, all is a pure mathematical fun.
Now we are going to infuse some physics into the above algebra. 

 In the scaling analysis (the perturbative renormalization 
group), we remove the high energy states progressively (renormalize)
to find the effective Hamiltonian which describes the 
same low temperature physics of the original Hamiltonian.
When the effective Hamiltonian does not change further
with renormalization (removal of the high energy states),
it is called the fixed point Hamiltonian. 
For the above general exchange interactions, the scaling 
equations up to the third order diagrams of Fig. \ref{scalemat} are 
(see the Appendices \ref{mgroup} \& \ref{3scale})
\bea
{\partial g^k \over \partial x}
 &=& {1\over 2} ~\left[ ~g^ig^j + g^jg^i \right] 
   - {1\over 4} ~ g^k ~ \mbox{Tr}\left[ g^ig^i + g^jg^j \right].
\eea
Here $(i,j,k)$ is cyclic. All our exchange interactions we are going to 
study satisfy the cyclic property such that Tr$[~]$ is independent of
the $x,y,z$ indices. Then we can see that the equations for the fixed points
have the same structure as in the $SU(2)$ spin algebra. That is, the 
exchange coupling matrices have the $SU(2)$ spin structure at the fixed 
points 
\bea
g^{i} &=& g^{*} ~ M^{i}, \\
g^{*} &=& {6 \over \mbox{Tr} \vec{M}^2 }, \\
\Lambda &=& {1\over 4} ~ \vec{M}^2.
\eea
The fixed point can be further specified by $\Lambda$ which measures
the conduction electron spin size. The number of channels is determined
by the ratio of the degeneracy of the relevant manifold and the spin size. 
Below we are going to illustrate this point explicitly.

 In passing we note that the same approach was independently applied 
to the metallic two-level system\cite{zarand}. 

\subsection{The one-channel $S_c=3/2, S_I=1/2$ exchange interaction}
 As shown in chapter II, the $\Gamma_8$ conduction electrons are 
coupled to the impurity pseudo spin $S_I=1/2$ with their pseudospin $S_c=3/2$ 
through the virtual fluctuations to the $f^2\Gamma_{4,5}$ triplet states.  

 For comparison, it is well understood that the one-channel $S_c=S_I=1/2$ 
exchange interactions ($\Gamma_6$ and $\Gamma_7$ conduction electrons) 
lead to the Fermi liquid strong coupling fixed point, while the two-channel
$S_c=S_I=1/2$ exchange interactions ($\Gamma_8$ conduction electrons) 
give rise to the non-Fermi liquid non-trivial fixed point. 

 The third order scaling equation is
\bea
{\partial g \over \partial x} 
 &=& g^2 - ~{1\over 3}~s_c(s_c+1)(2s_c+1)~g^3, ~~g=N(0)J. 
\eea
Here $s_c = 3/2$. We can find the stable fixed point $g^*=1/5$.

 We now argue that this new exchange interaction can lead to a non-Fermi 
liquid fixed point. In Fig.~\ref{screen}, we schematically show how 
the screening of the impurity spin can happen by the conduction electron
spin cloud. With the antiferromagnetic coupling, the energy is minimized
when two conduction electrons of $S^z_c=-3/2, -1/2$ are coupled to the 
impurity spin $S_I^{z}=1/2$. This picture does not violate the Pauli 
exclusion principle. Thus the effective impurity spin is $``S_I"=3/2$.
In the next shell screening, only the $S_c^{z}=3/2,1/2$ can come closer
to the effective impurity spin due to the Pauli exclusion principle. 
Hence the effective exchange interaction is antiferromagnetic. From this
argument, we can see that the strong coupling fixed point cannot remain 
stable. Since the weak coupling fixed point is unstable, as we shall show,
an intermediate fixed point must exist\cite{nozbland}. 
This is a non-Fermi liquid fixed point. We are separately 
studying this simple new exchange interaction 
using the numerical renormalization group(NRG) and conformal field theory
\cite{klctobe}. 

 We are next going to show that this new exchange interaction 
competes with the two-channel Kondo effect.

\subsection{The $\Gamma_8$ conduction electron space}
 In the $\Gamma_8$ conduction electron space, two different spins can be
realized as shown in section III. In summary, three different 
exchange interactions are possible via the $f^2 \Gamma_{3,4,5}$
\bea
H_1 (\Gamma_3)
 &=& J_3 ~ \sum_{n=\pm} \vec{S}_{cn} (0) \cdot \vec{S}_{\Gamma_7}
  = J_3 ~ \tau^0 \otimes \vec{S}_c (0) \cdot \vec{S}_{\Gamma_7}, \\
H_1 (\Gamma_4) 
 &=& J_4 ~ \vec{L}_{c} (0) \cdot \vec{S}_{\Gamma_7}
  = J_4 ~(~-\tau^0 - 2\tau^{i}~)~ \otimes S_c^{i} (0) ~ S_{\Gamma_7}^{i}, 
  \\
H_1 (\Gamma_5) 
 &=& J_5 ~ \vec{L}_{c}^{'} (0) \cdot \vec{S}_{\Gamma_7}
  = J_5 ~(~-\tau^0 + 2\tau^{i}~)~ \otimes S_c^{i} (0) ~ S_{\Gamma_7}^{i}. 
\eea
Here $\vec{S}_{\Gamma_7}$ is the impurity pseudo spin for the $f^1J=5/2~
\Gamma_7$ state. $\vec{L}_{c}(0)$ and $\vec{L}_{c}^{'}(0)$ are two different  
conduction electron pseudo spin representations for the same $S_c=3/2$ state 
manifold. Here we have rewritten the 
$S_c=3/2$ operators from a view point of the two-channel exchange 
interaction. 

All combined in the $\Gamma_8$ conduction electron space, we find
\bea
H_1 &=& J^i \otimes S_{c}^i ~S^i, \\
J^i &=& J_8 ~\tau^0 + J_{88} ~\tau^i, \\
J_8 &=& J_3 - J_4 - J_5, ~~J_{88} = J_5 - J_4. 
\eea
Physically, the $\tau^0$ term leads to the ordinary two-channel exchange  
interaction while the $\tau^i$ terms are channel-symmetry breaking and 
channel-mixing.
Then we can derive the scaling equation for the matrix ${\bf g}^{i} =
N(0)J^{i}$ in a compact way. 
In the $\Gamma_8$ conduction electron space, 
the third order scaling equations are
\bea
{\partial g_8 \over \partial x} 
 &=& g_8^2 - {1\over 2} ~g_{88}^2  - g_8~[~g_8^2 + g_{88}^2 ~], 
   ~~ g_8 = N(0)J_8 > 0, \\
{\partial g_{88} \over \partial x} 
 &=& - g_8 g_{88} - g_{88}~[~g_8^2 + g_{88}^2~], 
   ~~ g_{88} = N(0)J_{88}. 
\eea
We can identify the two-channel fixed point 
$(g_8^{*}, g_{88}^{*}) = (1,0)$ and two other interesting fixed 
points $(g_8^{*}, g_{88}^{*}) = (-1/5,\pm 2/5)$
\bea
{\bf g}^{*} &=& {1\over 5} ~ \vec{M}, \\
M^{i} &=& -\tau^0 \pm 2~\tau^{i}, \\
\Lambda &=& {3\over 2} \cdot {5\over 2} ~ \tau^0.
\eea
corresponding to the exchange interaction models generated by 
$f^2\Gamma_{4,5}$ in this restricted space, respectively. 
The linear analysis shows that all these fixed points are stable. 
Solving the above scaling equations, we can construct the flow diagram 
(see Fig.~\ref{g8flow}). There are three linear separatices along which
all the exchange interactions derived purely from the 
$f^2\Gamma_{3},\Gamma_4$, or $\Gamma_5$ CEF states lie. 
Along the $g_8$ axis, the two-channel $S_c=S_I=1/2$ 
exchange interaction due to virtual $f^2\Gamma_3$ charge fluctuations 
is located. 
Along the line $g_{88} = 2 g_8$, the one-channel $S_c=3/2, S_I=1/2$ 
exchange interaction due to virtual $f^2\Gamma_4$ charge 
fluctuations is located. 
Along the line $g_{88} = -2 g_8$, the one-channel $S_c=3/2, S_I=1/2$ 
exchange interaction due to virtual $f^2\Gamma_3$ charge 
fluctuations is located. 
>From this flow diagram, we can get the following qualitative results:
(1) When $2g_8 > |g_{88}|$, two-channel fixed point ground state is
realized;
(2) When $g_{88} > 0$ and $2g_8 < |g_{88}|$, the new fixed point
ground state (generated by $f^2\Gamma_5$) is realized.
(3) When $g_{88} < 0$ and $2g_8 < |g_{88}|$, the new fixed point
ground state (generated by $f^2\Gamma_4$) is realized.

 From a viewpoint of the two-channel exchange interaction, 
the above model Hamiltonian has channel asymmetry and 
exchange coupling anisotropy without the $\tau^1$ terms. From a 
well-known scaling argument\cite{nozbland,cragg,aniso}, 
the model will flow to the one-channel
fixed point (the channel with larger $z$ component of exchange coupling)
in the absence of the channel-mixing term. 
Though the $\tau^{3}$ term breaks channel symmetry, 
the additional term from $\tau^{1}$ mixes the two different 
channels. That is,
the channel asymmetry due to $\tau^{3}$ term is apparently washed out 
by the channel
mixing term restoring to the two-channel fixed point in the two-channel 
Kondo effect parameter regime. 

 We note that the $S_c=3/2$  fixed points have a net coupling strength 
(along the separatices) of $g^*=1/5$, safely in the perturbative regime
so that the scaling analysis is trustworthy.

\subsection{One-channel \& two-channel exchange interaction} 
 When we keep only the three atomic states $f^0, f^1J=5/2~\Gamma_7$, and 
$f^2\Gamma_3$, we find the one-channel ($\Gamma_7$) \& two-channel  
($\Gamma_8\pm$) Anderson hybridization interactions with 
the impurity pseudo spin $S_I=1/2$. 
This simple model shows competition between the Fermi liquid fixed point of
the one-channel Kondo effect and the non-Fermi liquid fixed point of 
the two-channel Kondo effect\cite{kimcox}. For completeness, we
will include the scaling analysis of this model here.
We studied this simplified model using the non-crossing approximation
\cite{kimcox}.

It can be deduced from the scaling theory that the low
temperature and the low energy physics is dominated by the one-channel 
or the two-channel Kondo effects depending on their relative
magnitude of the antiferromagnetic couplings. 
With the introduction of the exchange coupling matrix, 
it is more convenient for the derivation of the scaling equations
to rewrite the one-channel and two-channel exchange interactions 
in the following form
\bea
\tilde{H}_1 
 &=& {\bf J} \otimes \vec{S}_c (0) \cdot \vec{S}_I, \\
{\bf J} &=& \pmatrix{J_7 & 0 & 0 \cr
               0 & J_8 & 0 \cr 0 & 0 & J_8}. 
\eea
Here $\vec{S}_c$ and $\vec{S}_I$ are $S=1/2$ operators. 
The scaling equations of our simple model Hamiltonian up to 
the third order diagrams are 
\bea
{\partial {\bf g} \over \partial x} 
 &=& {\bf g}^2 - {1\over 2} ~ {\bf g} ~ \mbox{Tr} [{\bf g}^2] \cdots, 
   \\
{\bf g} &=& N(0) {\bf J}. 
\eea
The scaling equations in components are 
\bea
{\partial g_7 \over \partial x} 
 &=& g_7^2 - {1\over 2} ~g_7~[~g_7^2 + 2 g_8^2 ~], ~~ g_7 = N(0)J_7 > 0, \\
{\partial g_8 \over \partial x} 
 &=& g_8^2 - {1\over 2} ~g_8~[~g_7^2 + 2 g_8^2 ~], ~~ g_8 = N(0)J_8 > 0.
\eea
Here $x = \log(D/T)$. We can identify three fixed points related 
to one-, two-, and three-channel Kondo effects. 
The one-channel, strong coupling
fixed point $(g_7^*, g_8^*) = (\infty, 0)$ is beyond the perturbative 
regime, but is stable leading to the singlet ground state and Fermi liquid 
spectrum\cite{noz}. 
The three-channel fixed point $(2/3, 2/3)$ is 
stable along the line $g_7 = g_8$ in the $g_7-g_8$ plane, but unstable for 
any small perturbation from $g_7 = g_8$. Finally, the two-channel fixed point
$(0, 1)$ is stable leading to the logarithmically divergent thermodynamics
at zero temperature. 
>From the scaling analysis, we can infer the ground state physics: 
one-channel for $J_7 > J_8$; two-channel for $J_7 < J_8$; 
and three-channel for $J_7 = J_8$. 

\subsection{Complete $\Gamma_7\oplus\Gamma_8$ exchange coupling space}
 In the $J_c=5/2~\Gamma_7 \oplus \Gamma_8$ conduction 
electron subspace allowing for other than virtual $f^2\Gamma_3$ charge
fluctuations, all the possible exchange interactions can be 
written in a simple, compact form
\bea
H_1 &=& J^{i} \otimes S_c^{i} (0) ~ S_{\Gamma_7}^{i}, \\
J^{i} &=& J_7~\kappa^{0} + J_8~\tau^{0} + J_{78}~\kappa^{i} + 
     J_{88}~\tau^{i}.   
\eea
The matrices are defined in the section III. We already studied this 
model in some specific cases in the previous sections. 
We can find the scaling equations up to the third order 
(see the Appendices \ref{mgroup} \& \ref{3scale}) as
\bea
{\partial g_7 \over \partial x}
 &=& g_7^2 -{1\over 2}~g_{78}^2 
    - {1\over 2} ~g_7 ~[~g_7^2 + 2(g_{78}^2 + g_8^2 + g_{88}^2) ~], 
    \\
{\partial g_{78} \over \partial x}
 &=& -g_{78} \left[g_{88} + {1\over 2}(g_7+g_8)~)\right]
    - {1\over 2} ~g_{78} ~[~g_7^2 + 2(g_{78}^2 + g_8^2 + g_{88}^2) ~], 
    \\
{\partial g_8 \over \partial x}
 &=& g_8^2 - {1\over 2} g_{88}^2 - {1\over 4} g_{78}^2 
   - {1\over 2} ~g_8 ~[~g_7^2 + 2(g_{78}^2 +  g_8^2 + g_{88}^2) ~], 
   \\
{\partial g_{88} \over \partial x}
 &=& - g_8 g_{88} - {1\over 2} g_{78}^2 
   - {1\over 2} ~g_{88} ~[~g_7^2 + 2(g_{78}^2 +  g_8^2 + g_{88}^2) ~], 
  \\
 && g_7 = N(0) J_7, ~~ g_{78} = N(0) J_{78}, ~~g_8 = N(0) J_8, ~~
   g_{88} = N(0) J_{88}. 
\eea
From these scaling equations, a linearized analysis reveals that 
some of the fixed points we found in the previous sections remain stable. 
The relevant fixed points $(g_7^*, g_{78}^*, g_8^*, g_{88}^{*})$ are:\\
(i) $(\infty,0, 0,0)$: one-channel fermi liquid fixed point;\\
(ii) $(0,0,1,0)$: two-channel non-fermi liquid fixed point remains
stable against any small perturbation away from this fixed point
\bea
{\bf g}^{*} &=& M, ~~M^{i} = \tau^{0}, \\
\Lambda &=& {1\over 2} \cdot {3\over 2} ~ \tau^0. 
\eea
As discussed previously, $\Lambda$ measures the conduction pseudo-spin size
and the number of channels. Here the conduction electrons are described with  
the pseudo-spin $S_c=1/2$ and with two orbitally degenerate channels 
(Tr$\tau_0 =2$); \\
(iii) $(2/3,0,2/3,0)$: three-channel non-fermi liquid fixed point
\bea
{\bf g}^{*} &=& {2\over 3} ~M, \\
M^{i} &=& \kappa^0 + \tau^0, \\
\Lambda &=& {1\over 2} \cdot {3\over 2}~(~\kappa^0 + \tau^0~). 
\eea
is stable against the $\Gamma_7 - \Gamma_8$ mixing interaction or 
the $\Gamma_8 \pm$ channel mixing interaction, but unstable for any 
perturbation away from $g_7 = g_8$; \\
(iv) $(0,0,-1/5,\pm 2/5)$: $S_c=3/2$ fixed points are
\bea
{\bf g}_{\pm}^{i} 
 &=& {1\over 5} ~ M^{i}_{\pm}, \\
M^{i}_{\pm} &=& - \tau^0 \pm 2~\tau^{i}, \\
\Lambda &=& {3\over 2} \cdot {5\over 2} ~\tau^0. 
\eea
The ${\bf g}_{+}^{*}$ fixed point is stable, 
but the other $S_c=3/2$ fixed point ${\bf g}_{-}^{*}$
becomes unstable against the $\Gamma_7$ and 
$\Gamma_8$ mixing exchange interaction ($J_{78}$).
As we shall show below, the mixing interaction between $\Gamma_6$ and 
$\Gamma_8$ conduction electrons is relevant to the 
${\bf g}_{+}^{*}$ fixed point. 

There are another two unstable fixed points which we did not
consider before. The fixed points are 
$g_7^* = 2/11$, $g_8^* = -2/11$, $g_{78}^* = 0$, 
$g_{88}^* = \pm 4/11$
\bea
{\bf g}^{i} &=& {2\over 11} ~M^{i}, \\
M^{i} &=& \kappa^0 - \tau^0 \pm 2 ~ \tau^{i}, \\
\Lambda &=& {1\over 2} \cdot {3\over 2}~\kappa^0 
     + {3\over 2} \cdot {5\over 2} ~\tau^0.
\eea
At the fixed points the reduced exchange interaction is 
\bea
H_1 &=& {2\over 11}~\left( \vec{S}_c(0) + \vec{S}_c^{'} (0) \right)~ \cdot 
        \vec{S}_{\Gamma_7}. 
\eea
Here $S_c=1/2$ and $S_c^{'}=3/2$. That is, two different conduction 
``bands" with different spins are coupled to the impurity spin. 

 All the fixed points we considered so far have $g_{78}^{*}=0$. Since the 
$\Gamma_7 - \Gamma_8$ mixing interaction is  relevant perturbation
for some fixed point, we might imagine that some fixed points 
with a finite value of $g_{78}$ are possible. There are in total 
6 fixed points of which only two are stable 
\bea
A &:& g_7^* = -{2\over 9}, ~~g_8^* = -{2\over 9}, ~~
   g_{78}^* = \pm {4\sqrt{2} \over 9}, 
   ~~g_{88}^* = - {4\over 9}, \\
B &:& g_7^* = {1\over 3}, ~~g_8^* = -{1\over 6}, ~~
   g_{78}^* = \pm {\sqrt{2} \over 6}, ~~g_{88}^* = - {1\over 3}, \\
C &:& g_7^* = -{2\over 21}, ~~g_8^* = {2\over 15}, ~~
   g_{78}^* = \pm {8\sqrt{5} \over 105}, 
   ~~g_{88}^* = - {8\over 105}. 
\eea
Out of these, only the type $A$ fixed points are stable.
The type $B~\&~ C$ fixed points are unstable against 
a perturbation along one principal direction. 

The $A$ fixed points are
\bea
{\bf g}^{*}_{A} &=& {2\over 3} ~ M, \\
M^{i} &=& -{1\over 3} ~(~\kappa^0 + \tau^0~) 
     \pm {2\sqrt{2} \over 3} ~ \kappa^{i} - {2\over 3} ~ \tau^{i}, \\
\Lambda
 &=& {1\over 2} \cdot {3\over 2} ~(~\kappa^0 + \tau^0~).
\eea
The fixed points can be represented with the direct sum of three $S_c=1/2$
conduction electron pseudo spins. 
We further note that the complex coupling matrices can be simplified using 
the orthogonal transformation in the conduction electron channel space
\bea
&& M^{x} = \pmatrix{-1 & 0 & 0 \cr 0 & 0 & -1 \cr 0 & -1 & 0},~~
M^{y} = \pmatrix{-1 & 0 & 0 \cr 0 & 0 & 1 \cr 0 & 1 & 0},~~
M^{z} = \pmatrix{1 & 0 & 0 \cr 0 & -1 & 0 \cr 0 & 0 & -1}. \nonumber\\
&& 
\eea
After rearrangement and redefinition of the states with some
phase and another orthogonal transformation to diagonalize the 
spin-flipping terms in the channel space, we can rewrite the exchange 
interaction at the fixed points in a concise form
\bea
H_1 &=& J ~ \sum_{n=1}^{3} ~  \vec{S}_{cn} (0) \cdot \vec{S}_I.
\eea
This is the three-channel $S_c=S_I=1/2$ exchange interaction. 
We further note that the following exchange interactions
are all equivalent and result in the same physics, because
the $SU(2)$ algebra does not change (we call this the 
chiral -- ``handedness" -- symmetry).
\bea
H_1 &=& J ~\vec{S}_{c} (0) \cdot \vec{S}_I,  \\
H_1 &=& J ~(~- S_c^{x} S_I^{x} - S_c^{y} S_I^{y} + S_c^{z} S_I^{z}~), \\
H_1 &=& J ~(~ S_c^{x} S_I^{x} - S_c^{y} S_I^{y} - S_c^{z} S_I^{z} ~), \\
H_1 &=& J ~(~ - S_c^{x} S_I^{x} + S_c^{y} S_I^{y} - S_c^{z} S_I^{z} ~). 
\eea
We can show the equivalence between them by redefining the states 
with the introduction of additional phase.

 Type $B~\&~C$ fixed points can be written in a standard form
\bea
{\bf g}^{*}_{B} &=& {1\over 4} ~ M, \\
M^{i} &=& {4\over 3}~\kappa^0 - {2\over 3}~ \tau^0 
     \pm {2\sqrt{2} \over 3} ~ \kappa^{i} - {4\over 3} ~ \tau^{i}, \\
\Lambda
 &=& 1 \cdot 2 ~(~\kappa^0 + \tau^0~).
\eea
The spin representation is $S_c=1$.
\bea
{\bf g}^{*}_{C} &=& {2\over 35} ~ M, \\
M^{i} &=& - {5\over 3}~\kappa^0 + {7\over 3}~ \tau^0 
     \pm {4\sqrt{5} \over 3} ~ \kappa^{i} - {4\over 3} ~ \tau^{i}, \\
\Lambda
 &=& {5\over 2} \cdot {7\over 2} ~(~\kappa^0 + \tau^0~).
\eea
The spin representation is $S_c=5/2$.

 In the $\Gamma_6 \oplus \Gamma_8$ conduction electron space, we get the 
same physics as in the above analysis. The difference is that the 
$S_c=3/2$ fixed point stable against the $\Gamma_7 - \Gamma_8$ mixing
becomes unstable against the $\Gamma_6 - \Gamma_7$
conduction electron mixing. Hence both of the $S_c=3/2$ fixed points 
cannot remain stable in the presence of the mixing interactions between
the $\Gamma_8$ and the $\Gamma_{6,7}$ conduction electrons.

\subsection{The $\Gamma_6\oplus\Gamma_7\oplus\Gamma_8$ conduction electron 
space}
 In this section, we are going to study the full exchange interactions 
possible in our model approach. The extension of the scaling analysis to 
the full interactions is straightforward. 
\bea
H_1 &=& J^{i} \otimes S_c^{i} (0) ~ S_{\Gamma_7}^{i}, \\
J^{i} 
 &=& J_6~\gamma^{0} + J_7~\kappa^{0} + J_8~\tau^{0} 
    + J_{68}~\gamma^{i} + J_{78}~\kappa^{i} + J_{88}~\tau^{i}.   
\eea
Third order scaling equations are
\bea
{\partial g_6 \over \partial x}
 &=& g_6^2 -{1\over 2}~g_{68}^2 - {1\over 2} ~g_6 
   ~[~g_6^2+g_7^2 + 2(g_{68}^2+g_{78}^2 + g_8^2 + g_{88}^2) ~], \\
{\partial g_7 \over \partial x}
 &=& g_7^2 -{1\over 2}~g_{78}^2 - {1\over 2} ~g_7 
   ~[~g_6^2+g_7^2 + 2(g_{68}^2+g_{78}^2 + g_8^2 + g_{88}^2) ~], \\
{\partial g_8 \over \partial x}
 &=& g_8^2 - {1\over 2} g_{88}^2 - {1\over 4}~[~g_{68}^2+g_{78}^2~] 
   - {1\over 2} ~g_8 
   ~[~g_6^2+g_7^2 + 2(g_{68}^2+g_{78}^2 + g_8^2 + g_{88}^2) ~], \nonumber\\
 && \\
{\partial g_{68} \over \partial x}
 &=& g_{68} \left[g_{88} - {1\over 2}(g_6+g_8)~)\right] - {1\over 2} ~g_{68} 
   ~[~g_6^2+g_7^2 + 2(g_{68}^2+g_{78}^2 + g_8^2 + g_{88}^2) ~], \nonumber\\
 && \\
{\partial g_{78} \over \partial x}
 &=& -g_{78} \left[g_{88} + {1\over 2}(g_7+g_8)~)\right] - {1\over 2} ~g_{78} 
   ~[~g_6^2+g_7^2 + 2(g_{68}^2+g_{78}^2 + g_8^2 + g_{88}^2) ~], \nonumber\\
 && \\
{\partial g_{88} \over \partial x}
 &=& - g_8 g_{88} + {1\over 2}~[~g_{68}^2-g_{78}^2~] - {1\over 2} ~g_{88} 
   ~[~g_6^2+g_7^2 + 2(g_{68}^2+g_{78}^2 + g_8^2 + g_{88}^2) ~], \nonumber\\
 && \\
g_6 &=& N(0)J_6, ~~g_7 = N(0) J_7, ~~ g_{78} = N(0) J_{78}, ~~
   g_8 = N(0) J_8, ~~g_{88} = N(0) J_{88}. \nonumber\\
 && 
\eea
We can find various fixed points of the above scaling equations. We already
considered the cases (1) $g_{78}^*=g_{68}^*=0$; (2)$g_{78}^*\neq 0, 
g_{68}^*=0$;
(3) $g_{78}^*=0,g_{68}^*\neq 0$. One relevant question is to ask 
if the two-channel 
fixed point of the $\Gamma_8$ conduction electron manifolds 
remains stable in the presence of the full interactions. 
The two-channel fixed point turns out to be stable. 
There arises the four-channel fixed point due to the expansion of the 
conduction electron space
\bea
{\bf g}^{*} &=& {1\over 2} ~ M, \\
M^{i} &=& \gamma^0 + \kappa^0 + \tau^0, \\
\Lambda &=& {1\over 2} \cdot {3\over 2}~(~\gamma^0 + \kappa^0 + \tau^0~).
\eea
This fixed point as well as the three-channel fixed points remain stable 
against a small perturbation of the mixing interactions ($J_{68}, J_{78}$), 
but they are unstable away from $g_6 = g_7 = g_8$.
There is also an unstable two-channel fixed point when we fine tune 
$g_6=g_7$
\bea
{\bf g}^{*} &=& M, \\
M^{i} &=& \gamma^0 + \kappa^0, \\
\Lambda &=& {1\over 2} \cdot {3\over 2}~(~\gamma^0 + \kappa^0~).
\eea
This fixed point remains stable against the conduction electron 
mixing interactions but becomes unstable away from $g_6 = g_7$.

The two $S_c=3/2$ fixed points are unstable in the presence of the 
full mixing interactions between the $\Gamma_{6,7}$ and the $\Gamma_8$ 
conduction electrons. Since the $g_{68}$ and $g_{78}$ terms are 
relevant perturbations to the $S_c=3/2$ fixed points, we may 
imagine that these fixed points flow to the other fixed points 
with a finite value of mixing exchange couplings. We shall address
this question shortly. 

 Another relevant question is to ask if the new type $A$ 
three-channel fixed points discussed
in the previous section remain stable. These two fixed points 
which have 
\bea
{\bf g}^{*}_{A} &=& {2\over 3} ~ M, \\
M^{i} &=& -{1\over 3} ~(~\kappa^0 + \tau^0~) 
     \pm {2\sqrt{2} \over 3} ~ \kappa^{i} - {2\over 3} ~ \tau^{i}, \\
\Lambda
 &=& {1\over 2} \cdot {3\over 2}~(~\kappa^0 + \tau^0~)
\eea
are stable, modulo the separate equality of exchange constants in the
$\Gamma_6$ or $\Gamma_7$ space with those of the $\Gamma_8$ space. 

 Due to the expansion of the conduction electron space ($\Gamma_6$),
other fixed points related to the types $A, B, C$ (in the $\Gamma_7\oplus
\Gamma_8$ conduction electron space)
considered in the previous section arise
\bea
{\bf g}^{*}_{A} &=& {1\over 2} ~ M, \\
M^{i} &=& \gamma^0 -{1\over 3} ~(~\kappa^0 + \tau^0~) 
     \pm {2\sqrt{2} \over 3} ~ \kappa^{i} - {2\over 3} ~ \tau^{i}, \\
\Lambda
 &=& {1\over 2} \cdot {3\over 2} ~(~\gamma^0 + \kappa^0 + \tau^0~).
\eea
\bea
{\bf g}^{*}_{B} &=& {2\over 9} ~ M, \\
M^{i} &=& \gamma^0 + {4\over 3}~\kappa^0 - {2\over 3}~ \tau^0 
     \pm {2\sqrt{2} \over 3} ~ \kappa^{i} - {4\over 3} ~ \tau^{i}, \\
\Lambda
 &=& {1\over 2} \cdot {3\over 2} ~\gamma^0 + 1 \cdot 2 ~(~\kappa^0 + \tau^0~).
\eea
\bea
{\bf g}^{*}_{C} &=& {1\over 18} ~ M, \\
M^{i} &=& \gamma^0 - {5\over 3}~\kappa^0 + {7\over 3}~ \tau^0 
     \pm {4\sqrt{5} \over 3} ~ \kappa^{i} - {4\over 3} ~ \tau^{i}, \\
\Lambda
 &=& {1\over 2} \cdot {3\over 2} ~\gamma^0 
      + {5\over 2} \cdot {7\over 2} ~(~\kappa^0 + \tau^0~).
\eea
All these fixed points are unstable.

 We now consider the case: $g_{78}^*\neq 0,g_{68}^*\neq 0$.
There are 8 new fixed points which are not discussed in the previous
sections
\bea
{\bf g}^* &=& {1\over 42} ~ M, \\
M^{i} &=& 3~\gamma^0 - {7\over 3}~\kappa^0 - {7\over 3}~\tau^0
    \pm 2\sqrt{3} ~ \gamma^{i} \pm {2\sqrt{35} \over 3} ~ \kappa^{i} 
    + {4\over 3} ~ \tau^{i}, \\
M^{i} &=& -{7\over 3}~\gamma^0 + 3~\kappa^0 - {7\over 3}~\tau^0
    \pm {2\sqrt{35} \over 3} ~ \gamma^{i} \pm 2\sqrt{3} ~ \kappa^{i} 
    - {4\over 3} ~ \tau^{i}, \\
\Lambda &=& {7\over 2} \cdot {9\over 2} ~(~\gamma^0 + \kappa^0 + \tau^0~). 
\eea
These 8 fixed points are all unstable. The spin representation is 
$S_c=7/2$.

 In summary, various fixed points have been identified. See Table~\ref{fixpt}
for a list of all the possible fixed points found from the third order
scaling equations. The stable fixed points are: (i) one-channel fixed 
points; (ii) two-channel fixed point; (iii) three-channel
fixed point. We also identified a ``zoo" of unstable fixed points. 
At these unstable fixed points, various exchange interactions are 
possible: 
(iv) the higher conduction electron pseudo spins ($S_c=3/2,5/2,7/2$)
are interacting with the impurity pseudo spin $S_I=1/2$; 
(v) the multi-channel
conduction electrons with different size of pseudo spins interact with the 
impurity pseudo spin. The higher conduction electron pseudo spin models 
overscreen the impurity pseudo spin and thus can lead to non-trivial
non-Fermi liquid fixed points.

\section{Scaling analysis II: multiple conduction partial wave states.}
 Up to now, we have not included the multiple conduction electron 
partial wave states. By this we mean, e.g., additional $\Gamma_7$ 
states of odd spatial parity about the Ce$^{3+}$ site which may
mix with the principle $J=5/2~\Gamma_7$ states. 
In this section we are going to study the effect of them on various Kondo 
effects.
The point is that the multiple partial waves lead to the enhancement in the
exchange coupling of the original exchange interaction model. 

 In the cases of simple one-channel or two-channel exchange interactions, 
the algebra is very simple. When various exchange interactions are studied 
at the same time, the third order scaling equations become 
complex, large matrix differential
equations. As an example, we analyze the matrix scaling equations in the 
$\Gamma_8$ conduction electron space.

\subsection{Two-channel exchange interaction model.}
 We consider the two-channel exchange interaction with some arbitrary number 
($N$) of $\Gamma_8$ conduction electron states. Then the Anderson 
hybridization interaction reads in the two-channel sector
\bea
H_1 &=& \sum_{\epsilon} \sum_{i n\alpha} (-1)^{\alpha -1/2} V_i ~
    c_{\epsilon \Gamma_8^i n \alpha}^{\dagger} ~
    |f^1 \Gamma_7\bar{\alpha} > <f^2 \Gamma_3 n|
     + h.c.
\eea
We focus on one $f^2\Gamma_3$ CEF state for the moment.
Here the index $i$ labels the different $\Gamma_8$'s of the conduction 
electron partial waves projected at the atomic site. 
Then the Schrieffer-Wolff transformation leads to the effective 
exchange interaction
\bea
\tilde{H}_1 
 &=& \sum_{ij} \sum_{n=\pm}  ~ J_{ij} ~
    \vec{S}_{c\Gamma_8 ij n} (0) \cdot \vec{S}_{\Gamma_7}, \\
\vec{S}_{c\Gamma_8 ij n} (0)
 &=& \sum_{\epsilon \epsilon'} c_{\epsilon \Gamma_8^i n \alpha}^{\dagger} 
    ~ {1\over 2} \vec{\sigma}_{\alpha\beta} 
     c_{\epsilon' \Gamma_8^j n \beta}, \\
J_{ij} &=& \sqrt{J_i J_j}, ~~J_i = {2V_i^2 \over \epsilon_2 - \epsilon_1 }. 
\eea
Here $\epsilon_1 (\epsilon_2)$ is the energy of the atomic state 
$f^1\Gamma_7 (f^2\Gamma_3)$, respectively. Note that the orbital
degeneracy index $n$ in the $\Gamma_8$ manifold is not active in the
exchange interaction and thus provides the two degenerate channels. 
Except for the unphysical case of the same form of the projected 
conduction electron DOS for each $\Gamma_8$ symmetry, we can not 
diagonalize both the projected conduction electron Hamiltonian and 
the above interaction at the same time. 

 We demonstrate below how the physics can be simplified in this unrealistic, 
simple case: the same form of the projected conduction band DOS for 
each $\Gamma_8$ symmetry. 
$J_{ij} ~ \vec{S}_{c\Gamma_8 ij n}$ can be diagonalized 
in the $\Gamma_8$ label space with the orthogonal transformation. 
Since the conduction electron part is a unit matrix in the $\Gamma_8$ label
space, the diagonalization in the $\Gamma_8$ label space can be done 
at the same time for the conduction electron part and the exchange 
interaction part
\bea
\tilde{H}_1
 &=& J ~ \sum_{n=\pm} \vec{S}_{cn} (0) \cdot \vec{S}_{\Gamma_7}, \\
\vec{S}_{cn} (0) 
 &=& \sum_{\epsilon \alpha} \sum_{\epsilon' \beta} 
    c_{\epsilon n\alpha}^{\dagger} ~ {1\over 2} \vec{\sigma}_{\alpha\beta}
    c_{\epsilon' n\beta}, \\ 
c_{\epsilon \alpha} 
 &=&  \sum_{i} \sqrt{J_i\over J} ~ c_{\epsilon \Gamma_8^i n \alpha}, \\
J &=& \sum_{i} J_i.
\eea
We note that only one eigenvalue is non-zero and is given by $J$. All the 
other eigenvalues are identically zero. We have not included the other
transformations for the conduction electrons which are irrelevant. 
The other components of the transformed conduction electrons 
(nonbonding combinations) are decoupled from the impurity spin. 
We conclude that the multiple $\Gamma_8$ conduction electrons give rise to 
the enhanced two-channel exchange coupling with $J=\sum_i J_i$ and the 
nonbonding combinations decouple. 
 
We now consider more general case with all the excited $f^2\Gamma_3$ 
CEF states considered. In this case, the exchange coupling is symmetric
$J_{ij} = J_{ji}$ but without the relation $J_{ij} = \sqrt{J_iJ_j}$.
In general, the simplification
as above does not occur when we diagonalize 
$J_{ij} ~ \vec{S}_{c\Gamma_8 ij n}$. After diagonalization, we may enumerate
the eigenvalues of $J_{ij}$ in the descending order from the maximum 
positive value ($J_1$) to the smallest value ($J_N$). 
That is, $J_1 \geq J_2 \geq \cdots \geq J_N$. 
Then we can write the interaction in the diagonal form
\bea
\tilde{H}_1 
 &=& \sum_{i=1}^{N} \sum_{n=\pm} ~ J_i ~ \vec{S}_{c i n} (0) \cdot
    \vec{S}_{\Gamma_7}.
\eea
If there are any negative $J$'s, their magnitude will be renormalized
to zero at zero temperature. If $J_i$'s are nondegenerate, as generically 
expected,  
the model system will flow to the two-channel fixed point
for the most strongly coupled set of conduction electron states. 
If $J_1 = J_2 \neq J_3$, the system will flow to the four-channel 
fixed point, and so forth.

 Though we have already diagonalized the exchange couplings in the $\Gamma_8$
conduction electron label space for the same projected DOS, 
we can give the scaling argument for more general forms of both the symmetric 
exchange couplings and the projected DOS
\bea
\tilde{H}_1 
 &=& \sum_{ij} \sum_{n=\pm}  ~ J_{ij} ~
    \vec{S}_{cij n} (0) \cdot \vec{S}, \\
J_{ij} &=& J_{ji}, ~~ J_{ii} > 0.
\eea
The index $n$ labels the orbital degeneracy in the $\Gamma_8$ manifold
and the indices $i,j$ label the conduction electron partial waves. 
In the scaling analysis, it is more convenient to introduce the exchange 
coupling matrix $g_{ij} = \sqrt{N_i(0)N_j (0)} J_{ij}$. 
Here $N_i (\epsilon)$ is the projected DOS for the $i$-th
$\Gamma_8$ conduction electron. 
Then the third order scaling equations can be written in a very compact form
(${\bf g}$ is an $N\times N$ matrix)
\bea
{\partial {\bf g} \over \partial x} 
 &=& {\bf g}^2 - {\bf g} ~ \mbox{Tr} {\bf g}^2.
\eea
Now we can diagonalize the exchange coupling matrix and we assume it is 
already done. Then we can write the scaling equations for each 
diagonal component $G_i$
\bea
{\partial G_i \over \partial x} &=& G_i^2 - G_i ~ \sum_j G_j^2.
\eea
Whenever the diagonal elements are positive, they will grow initially,
e.g., with decreasing temperatures. However, the largest $G$ will determine
the low temperature physics. There are several fixed points of which only 
one fixed point is stable according to third order scaling analysis.
Third order scaling analysis says that the fixed point
$(G_1^*=1, G_2^*=G_3^{*}= \cdots =G_N^{*}= 0)$ is stable. 
This fixed point is none other
than the two-channel fixed point with the initial enhanced exchange 
couplings. 

In general, the $2M$-channel fixed point is $G_1^*=G_2^*= \cdots
=G_M^{*}=1/M$, $G_{i}^{*}=0$ for $i\geq M+1$ or
\bea
{\bf g}^{*}
 &=& {1\over M} ~ \pmatrix{ I_{M} & 0 \cr 0 & 0} 
\eea
In the fixed point coupling matrix, $I_M$ is an $M\times M$ unit matrix. 
This $2M$-channel fixed point has 
one eigenvalue of $-1/M$(irrelevant), 
$M-1$ eigenvalues of $1/M$ (relevant), 
and the remaining eigenvalues of $-1/M$ (irrelevant, 
corresponding to $G_i^{*}=0$ for $i\geq M+1$).  
This $2M$-channel fixed point is 
unstable along the $M-1$ principal directions and stable along the other 
remaining principal directions. 
Hence this $2M$-channel fixed point is stable only when
the bare couplings satisfy the relation: $G_1=\cdots=G_M>G_{i}$ for 
$i\geq M+1$. In the $M=1$ case, this fixed point becomes stable as mentioned 
above, since the unstable principal directions are absent. 

 In the case of the one-channel exchange interaction 
 models ($\Gamma_7$ and $\Gamma_6$
conduction electrons), we get the same results as in the two-channel
exchange interaction  
model case.

\subsection{Multiple $\Gamma_8$ conduction electron partial waves.} 
 Now we study the effect of multiple $\Gamma_8$ conduction partial waves 
on the Kondo effect in detail allowing for the $S_c=3/2$ fixed point
\bea
H_1 &=& \left[ {\bf J}_8 \otimes \tau^{0} 
     + {\bf J}_{88} \otimes \tau^{i} \right] \otimes 
     S_c^{i} (0) S^{i}. 
\eea
Then it is straightforward to find the scaling equations up to third order
(see the Appendices \ref{mgroup} \& \ref{3scale})
\bea
{\partial {\bf g}_8 \over \partial x} 
 &=& {\bf g}_8^2 - {1\over 2} ~ {\bf g}_{88}^2 
   - {\bf g}_8 ~ \mbox{Tr} \{ {\bf g}_{8}^2 + {\bf g}_{88}^2 \}, \\
{\partial {\bf g}_{88} \over \partial x} 
 &=& - {1\over 2} ~\left[ {\bf g}_{8} {\bf g}_{88} 
                         + {\bf g}_{88} {\bf g}_{8} \right] 
   - {\bf g}_{88} ~ \mbox{Tr} \{ {\bf g}_{8}^2 + {\bf g}_{88}^2 \}. 
\eea
Here ${\bf g}_{8}$ and ${\bf g}_{88}$ are reduced exchange coupling matrices. 
In general, when we consider the multiple conduction electron partial waves,
the exchange couplings become matrices. 

For definiteness, we first consider two $\Gamma_8$ conduction partial waves, 
say, coming from $J_c=5/2, 7/2$ multiplets. In this case, the exchange 
coupling matrices are $2\times 2$ matrices
\bea
{\bf J}_8 
 &=& \pmatrix{ J_{11} & J_{12} \cr J_{21} & J_{22} }; ~~
{\bf J}_{88} 
 = \pmatrix{ \tilde{J}_{11} & \tilde{J}_{12} \cr 
    \tilde{J}_{21} & \tilde{J}_{22} }. 
\eea
To solve the above scaling equations we may introduce the following 
decomposition of the exchange coupling matrices
\bea
{\bf g}_{8} &=& g_0~\rho^0 + g_1~\rho^1 + g_3~\rho^3, \\
{\bf g}_{88} 
 &=& \tilde{g}_0~\rho^0 + \tilde{g}_1~\rho^1 
    + \tilde{g}_3~\rho^3. 
\eea
Here $\rho^0$ is a $2\times 2$ unit matrix, and $\rho^{i}$'s are Pauli
matrices. These matrices are defined in the space of the conduction electron
partial waves where the $\Gamma_8$ manifolds are derived. 
Since the exchange coupling matrices are real and symmetric, 
there is no $\rho^2$ term. 
Then the scaling equations in components read
\bea
{\partial g_0 \over \partial x} 
 &=& |g|^2-{1\over 2} ~|\tilde{g}|^2 - 2g_0 ~(~|g|^2 + |\tilde{g}|^2~),
    \\
{\partial g_1 \over \partial x} 
 &=& 2g_0g_1 - \tilde{g}_0\tilde{g}_1 - 2g_1 ~(~|g|^2 + |\tilde{g}|^2~),
    \\
{\partial g_3 \over \partial x} 
 &=& 2g_0g_3 - \tilde{g}_0\tilde{g}_3 - 2g_3 ~(~|g|^2 + |\tilde{g}|^2~),
    \\
{\partial \tilde{g}_0 \over \partial x} 
 &=& - g\cdot \tilde{g} - 2\tilde{g}_0 ~(~|g|^2 + |\tilde{g}|^2~),
    \\
{\partial \tilde{g}_1 \over \partial x} 
 &=& -{1\over 2}~(~g_0\tilde{g}_1 + \tilde{g}_0g_1~)
    - 2\tilde{g}_1 ~(~|g|^2 + |\tilde{g}|^2~), \\
{\partial \tilde{g}_3 \over \partial x} 
 &=& -{1\over 2}~(~g_0\tilde{g}_3 + \tilde{g}_0g_3~)   
    - 2g_3 ~(~|g|^2 + |\tilde{g}|^2~), \\
|g|^2 &\equiv& g_0^2 + g_1^2 + g_3^2, \\
|\tilde{g}|^2 
  &\equiv& \tilde{g}_0^2 + \tilde{g}_1^2 + \tilde{g}_3^2, \\
g \cdot \tilde{g} 
 &\equiv& g_0 \tilde{g}_0 + g_1 \tilde{g}_1 + g_3\tilde{g}_3. 
\eea
There are several fixed points. First of all, the weak coupling fixed point
${\bf g}_{8}^* = {\bf g}_{88}^* = {\bf 0}$ is unstable. 
The four-channel fixed point 
\bea
&& {\bf g}_{8}^* 
 = {1\over 2} ~\pmatrix{1 & 0 \cr 0 & 1}, ~~
 {\bf g}_{88}^{*} = \pmatrix{0 & 0 \cr 0 & 0}
\eea
is stable along  $g_0$-axis, while unstable away from $g_1=g_3=0$. 
The two-channel fixed point is
\bea
{\bf g}_{8}^* 
  &=& {1\over 2} \left( \rho^0 + \sin \theta~\rho^1 + \cos\theta~\rho^3
       \right) 
  \to \pmatrix{1 & 0 \cr 0 & 0}.
\eea
Here $\theta$ is an arbitrary real number. We get the diagonal matrix along
the principal directions after an orthogonal transformation.
This fixed point is stable with enhanced two-channel
exchange coupling.

There is another class of the fixed points when ${\bf g}_{88}^* = \pm 2
{\bf g}_{8}^*$. The fixed point $g_0^*=-1/10, g_1^*=g_3^*=0$ or
\bea
{\bf g}_8^* &=& - {1\over 10}~ \pmatrix{1 & 0 \cr 0 & 1}
\eea
is unstable, which corresponds to the two-channel $S_c=3/2$ exchange
 interaction. 
The relevant stable fixed points are
$g_0^*=-1/5, |g_1^*|^2 + |g_3^*|^2 = 1/25$ or 
\bea
{\bf g}_8^* 
 &=& - {1\over 5} ~ \pmatrix{1+\cos\theta & \sin\theta \cr
                             \sin\theta & 1-\cos\theta}
  \to - {1\over 5}~ \pmatrix{1 & 0 \cr 0 & 0},
\eea
which are none other than 
new $S_c=3/2$ non-Fermi liquid fixed points with enhanced exchange 
coupling after the orthogonal transformation at the fixed points. 

 We can generalize the above analysis with arbitrary number of conduction
electron partial waves ($N$). We can determine the stability of the fixed
points relatively easily. The case ${\bf g}_{88}^*=0$ is equivalent to the
case of the two-channel exchange interaction with multiple partial waves. 
There are several multi-channel fixed points of which only the two-channel
fixed point is stable. In this case, the fixed points ${\bf g}_8^{*}$ are
solutions of 
\bea
&& {\bf g}_8^2 - \lambda ~ {\bf g}_8 =0, ~~ \lambda = \mbox{Tr} {\bf g}_8^2. 
\eea
The trivial weak coupling fixed point ${\bf g}_8^*=0$ is unstable. To find
non-trivial fixed points, we introduce the orthogonal transformation to 
diagonalize the above matrix equation. Then we find the general
$2M$-channel fixed points (after the orthogonal transformation)
\bea
{\bf g}_8^* 
 &=& {1\over M} ~ \pmatrix{I_M & 0 \cr 0 & 0}, ~~
 {\bf g}_{88}^{*} = 0
\eea
with $\lambda = 1/M$. Here $I_M$ is an $M\times M$ unit matrix. $M$ can have
any value of $M=1,2,\cdots, N$.
Now we consider the stability of the fixed points. We will show that only
the fixed point for $M=1$ is stable.
Close to the fixed points, we can linearize the scaling
equations to find 
\bea
{\partial \delta{\bf g}_8 \over \partial x} 
 &=&  {\bf g}_8^* \delta {\bf g}_8 + \delta {\bf g}_8 ~ {\bf g}_8^{*} 
   - \lambda ~ \delta {\bf g}_8 - {\bf g}_8^{*} ~ 
    \mbox{Tr} \{ {\bf g}_8^{*} \delta {\bf g}_8 
                + \delta {\bf g}_8 ~ {\bf g}_8^{*}\}, \\
{\partial \delta {\bf g}_{88} \over \partial x} 
 &=& - {1\over 2} ~\left[ {\bf g}_{8}^{*} \delta {\bf g}_{88} 
                         + \delta {\bf g}_{88} {\bf g}_{8}^{*} \right] 
   - \lambda ~\delta {\bf g}_{88}, \\ 
\lambda 
 &=& \mbox{Tr} \{{\bf g}_{8}^{*} {\bf g}_{8}^{*}\} = {1\over M}. 
\eea
Diagonalization of the first equation can be done by the same orthogonal 
transformation introduced for finding the fixed points. Here it is enough
to consider a small perturbation along the principal directions, since
any other forms of perturbation can be written as a linear combination of 
those along the principal directions. Then along the principal directions,
we find
\bea
{\partial \delta g_8^{i} \over \partial x} 
 &=& {1\over M} ~\delta g_8^{i} 
   - {2\over M^2}~\sum_{j=1}^{M} \delta g_8^{j}, ~~i=1,2, \cdots, M, \\ 
{\partial \delta g_8^{i} \over \partial x} 
 &=& - {1\over M} ~\delta g_8^{i}, ~~i=M+1,M+2, \cdots, N.  
\eea
Thus this $2M$-channel fixed point is stable along the $N-M$ principal
directions (corresponding to $i=M+1,M+2,\cdots,N$). And it is 
stable along one direction and unstable along the other $M-1$ directions 
after the diagonalization for $i=1,2,\cdots,M$. For a small perturbation
along the principal directions of ${\bf g}_{88}^*=0$ (in fact, those of 
${\bf g}_8^{*}$),
we find
\bea
{\partial \delta g_{88}^{i} \over \partial x} 
 &=& - {2\over M} ~\delta g_{88}^{i}, ~~i=1,2, \cdots, M, \\ 
{\partial \delta g_{88}^{i} \over \partial x} 
 &=& - {1\over M} ~\delta g_{88}^{i}, ~~i=M+1,M+2, \cdots, N.  
\eea
Thus the $2M$-channel fixed point is stable for any small perturbation of
${\bf g}_{88}$. We conclude that only the two-channel fixed point for $M=1$
is stable when we consider the multiple conduction electron partial 
waves.  

 Another class of fixed points ${\bf g}_{88}^{*} = \pm 2 {\bf g}_8^{*}$ 
are analyzed in the same way. 
The identification of the fixed points are simple: there
are $M$-channel $S_c=3/2$ fixed points of which only the case $M=1$ is 
stable. The $M$-channel $S_c=3/2$ fixed point matrix is
(after diagonalization) 
\bea
{\bf g}_{88}^{*} &=& \pm 2 {\bf g}_8^{*}, \\
{\bf g}_8^{*}
 &=& - {1\over 5M} ~ \pmatrix{I_M & 0 \cr 0 & 0}. 
\eea
Here $I_M$ is an $M\times M$ unit matrix.
Now we study the stability of this fixed point using the linearized
scaling equations close to the fixed point
\bea
{\partial \delta{\bf g}_8 \over \partial x} 
 &=& {\bf g}_8^* \delta {\bf g}_8 + \delta {\bf g}_8~{\bf g}_8^{*}
   - {1\over 2} \left[ {\bf g}_{88}^* \delta {\bf g}_{88}
        + \delta {\bf g}_{88}~ {\bf g}_{88}^{*} \right] 
   - \lambda ~ \delta {\bf g}_8 \nonumber \\
 && -  {\bf g}_8^{*} ~ 
   \mbox{Tr} \{ {\bf g}_8^{*} \delta {\bf g}_8 
                + \delta {\bf g}_8~ {\bf g}_8^{*} 
                + {\bf g}_{88}^{*} \delta {\bf g}_{88} 
                + \delta {\bf g}_{88}~ {\bf g}_{88}^{*} \}, \\
{\partial \delta {\bf g}_{88} \over \partial x} 
 &=& - {1\over 2} ~\left[ {\bf g}_{88}^{*} \delta {\bf g}_{8} 
                         + \delta {\bf g}_{8} {\bf g}_{88}^{*}
                         + {\bf g}_{8}^{*} \delta {\bf g}_{88} 
                         + \delta {\bf g}_{88} {\bf g}_{8}^{*} \right] 
   - \lambda ~\delta {\bf g}_{88} \nonumber\\
 && - {\bf g}_{88}^{*} ~
   \mbox{Tr} \{ {\bf g}_8^{*} \delta {\bf g}_8 
                + \delta {\bf g}_8~ {\bf g}_8^{*} 
                + {\bf g}_{88}^{*} \delta {\bf g}_{88} 
                + \delta {\bf g}_{88}~ {\bf g}_{88}^{*} \}, \\
\lambda 
 &=& \mbox{Tr} \{ {\bf g}_{8}^{*}{\bf g}_{8}^{*} 
   + {\bf g}_{88}^{*}{\bf g}_{88}^{*} \} = {1\over 5M}. 
\eea 
Again it is enough to consider the linear stability along the principal 
directions. Then we find for ${\bf g}_{88}^{*} = 2{\bf g}_{8}^{*}$
\bea
{\partial \delta g_8^{i} \over \partial x} 
 &=& - {3\over 5M} ~ \delta g_{8}^{i} + {2\over 5M} ~ \delta g_{88}^{i}
   - {2\over 25M^2} ~\sum_{j=1}^{M} 
       ~(~\delta g_8^{j} + 2\delta g_{88}^{j} ~), ~~i=1,2, \cdots, M, 
     \nonumber\\
 && \\
{\partial \delta g_{88}^{i} \over \partial x} 
 &=& {2\over 5M} ~ \delta g_{8}^{i} 
   - {4\over 25M^2} ~\sum_{j=1}^{M} 
       ~(~\delta g_8^{j} + 2\delta g_{88}^{j} ~), ~~i=1,2, \cdots, M, \\
{\partial \delta g_8^{i} \over \partial x} 
 &=& - {1\over 5M} ~ \delta g_{8}^{i}, ~~i=M+1,M+2, \cdots, N, \\
{\partial \delta g_{88}^{i} \over \partial x} 
 &=& - {1\over 5M} ~ \delta g_{88}^{i}, ~~i=M+1,M+2, \cdots, N. 
\eea
As we can see, the fixed point is stable for any small perturbation
along the principal directions with $i=M+1,M+2, \cdots, N$. For a small  
perturbation along the principal directions of $i=1,2,\cdots,M$, we can 
rewrite the linearized scaling equations as
\bea
{\partial \delta g_8^{i} \over \partial x} 
 &=& \left( -{3\over 5M}~\delta_{ij} - {2\over 25M^2}~A_{ij} \right)~
     \delta g_8^{j} 
  + \left( {2\over 5M}~\delta_{ij} - {4\over 25M^2}~A_{ij} \right)~
     \delta g_{88}^{j}, \nonumber\\
 && \\
{\partial \delta g_{88}^{i} \over \partial x} 
 &=& \left( {2\over 5M}~\delta_{ij} - {4\over 25M^2}~A_{ij} \right)~
     \delta g_8^{j} 
  - {8\over 25M^2}~A_{ij}~\delta g_{88}^{j}.
\eea
Here $\delta_{ij}$ is the Kronecker delta and $A_{ij} = 1$ for all 
$i,j=1,2,\cdots,M$. We further note that the combinations
$\delta g \equiv \delta g_8^i + 2~\delta g_{88}^i$ and 
$\delta \tilde{g} \equiv 2~\delta g_8^i - \delta g_{88}^i$ have 
simple scaling equations
\bea
{\partial \delta g^{i} \over \partial x} 
 &=& \left( {1\over 5M}~\delta_{ij} - {2\over 25M^2}~A_{ij} \right)~
     \delta g^{j}, \\ 
{\partial \delta \tilde{g}^{i} \over \partial x} 
 &=&  - {2\over 25M^2}~\delta \tilde{g}^{i}.  
\eea
Note that the diagonalization of the combination $g$ leads to 
one eigenvalue of $-1/M$ and $M-1$ eigenvalues of $1/M$.
The combination $\tilde{g}$ is an irrelevant perturbation, while the
combination $g$ is an relevant perturbation for $M>1$. 
Hence we conclude that only the one-channel $S_c=3/2$ fixed point is 
stable, while the $M$-channel $S_c=3/2$ fixed point is unstable for $M>1$.

 Now we are going to close this section after discussing how to extend 
our analysis to the full interaction case. In this case, the exchange 
interaction is a little bit complicated
\bea
H_1 &=& J^i \otimes S_c^{i} ~ S_{\Gamma_7}^{i}, \\
J^{i}
 &=& J_6\otimes\gamma^{0} + J_7\otimes\kappa^{0} + J_8\otimes\tau^{0}
     + J_{68}\otimes\gamma^{i} + J_{78}\otimes\kappa^{i}
     + J_{88}\otimes\tau^{i}.
\eea
Here $J_i$'s are matrices whose dimension is determined by the 
number of the conduction electron partial waves. 
Applying the third order scaling equations, we find more
fixed points. However, only the stable fixed points we considered
in the previous section -- one-, two-, and three-channel fixed points --
remain stable in the presence of the multiple conduction electron 
partial waves. As we found in the analysis of Sec. IV, 
a big zoo of new unstable
multi-channel fixed points are generated. 
     
 Our overall conclusion is that the one-, two-, and three-channel 
fixed points remain stable 
with the enhanced bare exchange couplings, even in the presence of the
multiple conduction electron partial waves.

\section{Discussion and Conclusion.}
 We have introduced and studied a realistic model Hamiltonian for Ce$^{3+}$
impurities with three configurations ($f^0$, $f^1$, $f^2$), which 
are embedded in cubic normal metals. Using the third order scaling theory,
we analyzed a various terms in our model Hamiltonian. 

In our study, a new exchange interaction has been discovered, 
which may lead to the non-Fermi liquid $S_c=3/2$ fixed point. 
This model is under study using the numerical renormalization group
(NRG) and conformal field theory\cite{klctobe}. Unfortunately, this
new fixed point becomes unstable against the mixing interactions
between the $\Gamma_8$ and the $\Gamma_{6,7}$ conduction electrons.

 Within our model study, we found that the one-channel (the $\Gamma_6$ or 
the $\Gamma_7$ conduction electron manifold), the two-channel
(the $\Gamma_8$ conduction electron manifold),
and the three-channel (the $\Gamma_6 \oplus \Gamma_8$ or the 
$\Gamma_7 \oplus \Gamma_8$ manifold) 
$S_c=S_I=1/2$ non-Fermi liquid fixed points are stable.
From the scaling equations, we also identified several unstable
fixed points. Of special interest are 
the fixed points with the higher conduction 
electron spins $S_c=3/2, 5/2, 7/2$, which may generate
non-Fermi liquid physics.  The multi-channel exchange interactions with
the different conduction electron spin size are also identified at 
the unstable fixed points. We may ask what the physical meaning of 
these unstable fixed points are. When some specfic exchange interaction
coupling is strong enough, the system will show the physics 
at high temperatures
relevant to this exchange interaction model before flowing finally
to the stable fixed point at zero temperature.

 The multiple conduction electron partial waves generate 
many unstable multi-channel fixed points.   Again we find 
stability of the one-channel, two-channel and three-channel 
fixed points.  

We close by noting a slight discrepancy between our work and that
of Koga and Shiba\cite{japan}, who studied the possible non-Fermi liquid 
ground states of an $f^3,J=9/2,\Gamma_6$ doublet interacting with
conduction electrons.   The exchange coupling with conduction electrons
is mediated by virtual $f^3-f^2$ charge fluctuations.  
In an NRG study of this model with a particular choice of coupling
strengths, it was found that a fermi liquid ground state always 
arose.  However, while the initially derived hamiltonian is identical
in form to that considered in our paper for the 
$\Gamma_6\oplus\Gamma_7\oplus\Gamma_8$ sector of conduction states, 
we find that the NRG Hamiltonian used in this paper has explicit
broken channel symmetry for the effective two-channel coupling.  
We believe that this may be the source of the divergent conclusions
between our work and theirs. 

\acknowledgments

This research was supported by a grant
from the U.S. Department of Energy, Office of Basic Energy Sciences,
Division of Materials
Research.  We thank L. N. Oliveira, and J.W. Wilkins for
stimulating interactions.

\appendix

\section{Mixing matrix elements}
\label{mixing}

 In this Appendix, we shall discuss briefly how to project the atomic 
electron operators to the atomic electron Fock space. We will use the 
Wigner-Eckart theorem in the projection and calculate the fractional
parentage coefficients. 

 To treat the strong on-site Coulomb interaction accurately, we introduce 
the projected states belonging to the atomic electron Fock space. In our 
model for Ce$^{3+}$, we assume the strong on-site Coulomb interaction, 
strong spin-orbit coupling, and strong crystalline electric field (CEF) 
in the atomic electron space. 
We project the atomic electron 
Fock space to the three configurations $f^0, f^1, f^2$ which are relevant
to the trivalent Ce impurities embedded in normal metals.
The total orbital angular momentum ($\vec{L}$) and the total spin angular 
momentum ($\vec{S}$) operators are good quantum numbers only in the presence
of the Coulomb interaction. When we include the LS coupling as well as the
Coulomb interaction, now the total angular momentum operator, 
$\vec{J} = \vec{L} + \vec{S}$, is a good quantum number. 
Below we show the projection of the atomic electrons when the total 
angular momentum operator $\vec{J}$ commutes with the atomic Hamiltonian. 
The extension to the crystal electric field case is straightforward. 

 The projection of the atomic electron operators to the atomic electron 
Fock space is realized by
\bea
{\bf 1} 
 &=& |f^0><f^0| + \sum_{i} |f^1; i><f^1; i| + \sum_{j} |f^2; j><f^2; j|,
  \\
f_{\mu} &\to & {\bf 1} f_{\mu} {\bf 1}.
\eea
Here the right hand side of ${\bf 1}$ is the restricted completeness 
relation to the three configurations $f^0, f^1, f^2$. 
The labels $i,j$ count all the possible states in the $f^1$ and
$f^2$ configurations, respectively. 

 In the $f^1$ configuration, the atomic electron states are labeled by 
$j_1 = 5/2, 7/2$ and their azimuthal quantum numbers. That is,
\bea
 && |f^1 {5\over 2} \mu_1> ~~\mbox{and} ~~ |f^1 {7\over 2} \mu_1>. \nonumber
\eea
In the $f^2$ configuration, the total angular momentum ranges from $0$ 
through $6$. The atomic electron states can be represented by
$|f^2 LSJ\mu_2>$. For the spin singlet states, $J=L=0,2,4,6$. For the spin 
triplet states, $L=1,3,5$ and $J=L,L\pm 1$. 
We are now in a position to find the projected atomic electron operator
of $f_{j_c\mu_c}$.
\bea
f_{j_c\mu_c}
 &\to& |f^0><f^0|~f_{j_c\mu_c}~\sum_{j_1\mu_1} |f^1 j_1\mu_1><f^1 j_1\mu_1|
    \nonumber\\
 && + \sum_{j_1\mu_1} |f^1 j_1\mu_1><f^1 j_1\mu_1| ~f_{j_c\mu_c}~
    \sum_{LSJ\mu_2} |f^2 LSJ\mu_2><f^2 LSJ\mu_2| \nonumber\\ 
 &=& |f^0><f^1j_c\mu_c| 
   + \sum_{j_1\mu_1} \sum_{LSJ\mu_2} 
   \Lambda (j_c\mu_c; f^1j_1\mu_1| f^2 LSJ\mu_2) ~ |f^1 j_1\mu_1> <f^2 LSJ\mu_2|, 
   \nonumber\\
 && 
\eea
\bea
\Lambda (j_c\mu_c; f^1j_1\mu_1| f^2 LSJ\mu_2)
 &\equiv& <f^1 j_1\mu_1| ~f_{j_c\mu_c}~|f^2 LSJ\mu_2>. 
\eea
Here $\Lambda$ measures the mixing strength between $f^1$ and $f^2$ 
atomic states. Applying the Wigner-Eckart theorem, we can rewrite the 
mixing strength $\Lambda$ as the Clebsch-Gordan coefficients multiplied
by a prefactor 
\bea
\Lambda (j_c\mu_c; f^1j_1\mu_1| f^2 LSJ\mu_2)
 &=& K(j_c;f^1j_1|f^2LSJ) ~ <j_c\mu_c; j_1\mu_1|J\mu_2>.  
\eea
Here the prefactor $K(j_c;f^1j_1|f^2LSJ)$ is the fractional parentage 
coefficient and is listed in Tables \ref{fracs},\ref{fract}.

 In the strong LS coupling limit, we now illustrate how we can calculate 
the matrix elements $< f^1 j_1 \mu_1 ~|~f_{j_c \mu_c}~|~f^2LS: J \mu_2 >$ 
to find the fractional parentage coefficients. 
We construct symmetry eigenstates of the total angular momentum in the $f^2$
configuration. 
\bea
&& |f^2LS=1:J\mu_2> \nonumber\\
 && \hspace{1.0cm} = \sum_M \sum_{m_1 m_2} <3m_1; 3m_2|LM> \nonumber\\
 && \hspace{1.0cm} \times {1\over \sqrt{2}}~
   \left( 
    \ba{l}  
     <11; LM|J\mu_2 > ~ f_{m_1\up}^{\dagger} f_{m_2\up}^{\dagger} |0> \\ 
     + <10; LM|J\mu_2 > ~ {1\over \sqrt{2}}
       ~(~f_{m_1\up}^{\dagger} f_{m_2\down}^{\dagger}
          + f_{m_1\down}^{\dagger} f_{m_2\up}^{\dagger} ~)~|0> \\
     + <1\bar{1}; LM|J\mu_2 > ~
       f_{m_1\down}^{\dagger} f_{m_2\down}^{\dagger} |0> 
    \ea 
   \right), \\
&& |f^2LS=0:J\mu_2> \nonumber\\
 && \hspace{1.0cm} = \sum_M \sum_{m_1 m_2} <3m_1; 3m_2|LM> \nonumber\\
 && \hspace{1.0cm} \times <00; LM|J\mu_2 > ~ {1\over 2 }
    ~(~f_{m_1\up}^{\dagger} f_{m_2\down}^{\dagger}
      - f_{m_1\down}^{\dagger} f_{m_2\up}^{\dagger} ~)~|0>. 
\eea
These symmetry states are not energy eigenstates of the atomic
Hamiltonian (Coulomb interaction and LS coupling) in the $f^2$ configuration. 
Mixing can occur among the same $J$ multiplets, but this is neglected
in the LS limit. 
Below we evaluate the desired matrix elements for the above symmetry 
eigenstates.

{\it Spin singlet case:} In this case, $J=L$ and only {\bf even} $L=J$ terms 
are nonvanishing.
\bea
&& < f^1 j_1 \mu_1 ~|~f_{j_c \mu_c}~|~f^2LS=0: J \mu_2 > \nonumber\\
 && \hspace{0.5cm} = 
  \sum_{\alpha_c m_c m_1} ~(-1)^{1/2-\alpha_c} ~ 
    < j_c\mu_c | {1\over 2}\alpha_c;3m_c > ~
    <j_1\mu_1| {1\over 2}\bar{\alpha}_c;3m_1 >~
    <J\mu_2 | 3m_c;3m_1> \nonumber\\
 && \hspace{0.5cm} =
  (-1)^{J+j_c+3/2} ~ \sqrt{(2j_c+1)(2j_1+1)} ~ 
  \left\{ \ba{ccc} j_c & j_1 & J \\ 3 & 3 & 1/2 \ea \right\} ~
  <J\mu_2 | j_c \mu_c; j_1 \mu_1>. 
\eea
The final result is proportional to the correct Clebsch-Gordan coefficients.
Here we introduced the $6-j$ symbols\cite{condon}. The explicit calculation 
gives the fractional parentage coefficients in terms of $6-j$ symbols. 
The fractional parentage coefficients are listed in Table \ref{fracs}. 

{\it Spin triplet case:} In this case, only {\bf odd} $L$ terms are 
nonvanishing.
\bea
&& < f^1 j_1 \mu_1 ~|~f_{j_c \mu_c}~|~f^2LS=1: J \mu_2 > \nonumber\\
 && \hspace{0.5cm} = 
 \sqrt{2} ~\sum_{\alpha_1\alpha_c} \sum_{Mm} \sum_{m_1m_c} ~
   <LM|3m_c; 3m_1> ~ <1m |{1\over 2}\alpha_c; {1\over 2}\alpha_1>
    \nonumber\\
 && \hspace{0.5cm} \times 
    < j_c\mu_c | {1\over 2}\alpha_c; 3m_c> ~
    < j_1\mu_1 | {1\over 2}\alpha_1; 3m_1> ~ <1m; LM| J\mu_2 > \\
 && \hspace{0.5cm} = K(j_c; f^1j_1~|~f^2 LS=1:J)~
   <J\mu_2 | j_c \mu_c; j_1 \mu_1>, \\
&& K(j_c; f^1j_1~|~f^2 LS=1:J) \nonumber\\ 
 && \hspace{0.5cm} = 
   \sqrt{6(2L+1)(2j_1+1)(2j_c+1)} \nonumber\\
 && \hspace{0.5cm} \times 
   \sum_{Q} (2Q+1) ~ (-1)^P ~ 
      \left\{ \ba{ccc} 3 & 3 & L \\ J & 1 & Q \ea \right\} ~
      \left\{ \ba{ccc} 3 & j_c & 1/2 \\ 1/2 & 1 & Q \ea \right\} ~
      \left\{ \ba{ccc} J & j_c & j_1 \\ 1/2 & 3 & Q \ea \right\}.
     \nonumber\\
 && 
\eea
Here $P=2j_c+2L+2j_2 +1$.
For a given $L$, there are three values of $J=L-1, L, L+1$.
The fractional parentage coefficients are listed in 
Table \ref{fract}.

\section{Multiplicative renormalization}
\label{mgroup}

 When we apply the perturbation theory to the Kondo problems, logarithmic
divergences appear at all orders in the coupling. 
This infrared divergence can be handled by
renormalizing the exchange coupling constant for the interaction vertex, and
renormalizing the ``mass enhancement" parameter for the pseudofermion Green's
function. Then we do the perturbation theory in the renormalized parameters
in the hope that it leads to converging solutions.

 Though the renormalization can be defined in several different ways,
we may consider the progressive removal of the conduction band edge states 
for definiteness. At temperature $T$, only the conduction electrons 
(thermally excited)
inside the band of order $T$ with respect to the Fermi level play an 
important role in determining physical properties. 
Thus we can integrate out the band edge states (virtually excited states) 
to find the effective Hamiltonian. When the renormalized Hamiltonian is
of the same form as the original one, we can find the scaling equations 
for the model parameters. Though the following analysis is
restricted to the perturbative regime,
we can derive qualitative results out of this. For 
quantitative results, the full numerical renormalization group
method is required.

 Here we just sketch how to generalize the multiplicative renormalization 
group\cite{scale2,scale3} to the following form of the exchange interaction.
\bea
H_1 &=& \sum_{i} J^{i}\otimes S_c^{i} (0) ~ S_I^{i}. 
\eea
Here $S_I=1/2$ is the impurity spin, and $S_c=1/2$ is the conduction electron
spin. The exchange coupling is arbitrary square matrix. 
For this exchange interaction, the interaction vertex is
\bea
\Gamma^{a}_{pq\alpha\beta;\mu\nu} 
 &=& J_{pr}^{a} ~ \tilde{\Gamma}_{rq}^{a} ~ S_{c\alpha\beta}^{a} ~
   {1\over 2} ~ \sigma_{\mu\nu}^{a}. 
\eea
This exchange interaction describes the most general interaction introduced
in chapter II. Though the Schrieffer-Wolff transformation generates the 
potential scattering terms, we are not going to include them here. 

 In the Kondo problem, the multiplicative renormalization is defined 
as\cite{scale2,scale3} 
\bea
&& G \rightarrow z_1 G, \\
&& F \rightarrow z_2 F, \\
&& \tilde{\Gamma}^{a} \rightarrow [Z^{a}]^{-1} ~ \tilde{\Gamma}^{a}, 
  \\
&& J^{a} \rightarrow {1 \over z_1 z_2} ~ J^{a}Z^{a}. 
\eea
Here $G$ and $F$ are the Green's functions for the conduction electrons 
and the pseudo fermions representing the local spins. 
$\tilde{\Gamma}$ is the scattering interaction vertex (matrix), 
and $J$ is the exchange coupling (matrix). Since the renormalization of 
the conduction electron is negligible, $z_1 \to 1$. Here we note that 
$Z^{a}$ ia a matrix and $a=x,y,z$. 

After following the same algebraic steps as in Ref\cite{scale2,scale3}, we find
the following scaling equations. 
\bea
{\partial J_{\rm inv}^{a} \over \partial x}  
 &=& f_d (J_{\rm inv})~J_{\rm inv}^{a} + 
    J_{\rm inv}^{a} ~ F_{\tilde{\Gamma}}^{a} (J_{\rm inv}). 
\eea
The functions $f_d$ and $F_{\tilde{\Gamma}}^{a}$ (matrix) are 
found from the irreducible or skeleton digrams of 
$d(\omega)$ and $\Gamma (\omega)$ using the perturbation 
theory. 
\bea
d(\omega) &=& 1 + f_d (J) ~ K (\omega), \\
\tilde{\Gamma}^{a} (\omega) 
 &=& 1 + F^{a}_{\tilde{\Gamma}} (J) ~ K(\omega), \\
K (\omega) 
 &=& \log \left|{D\over \omega}\right| + i\pi \theta (\omega). 
\eea

\section{Scaling equations up to third order}
\label{3scale}

 We consider the following form of the interaction.
\bea
\tilde{H}_1 
 &=& \sum_{i} {\bf J}^{i} \otimes S_c^{i} (0) ~ S^i_{I}. 
\eea
Here $S^i$ is the $i$-th component of the impurity spin operator ($S_I=1/2$) 
and $S_c^i (0)$ is the $i$-th component of the conduction electron spin 
operator ($S_c=1/2$) projected at the impurity site. 
${\bf J}^{i}$ is the exchange coupling matrix.

 It is straightforward to derive the scaling equations. 
The second order diagrams give
\bea
&& -N(0)K(\omega) ~S^j S^i ~J^j\otimes S_c^j J^i \otimes S_c^i \\
&& N(0)K(\omega) ~S^i S^j ~J^j\otimes S_c^j J^i \otimes S_c^i 
\eea 
The first comes from the particle excitation, while the second from 
the hole excitation. Their contribution is
\bea
&& N(0)K(\omega)~ i\epsilon_{ijk}S^k ~J^j J^i \otimes S_c^jS_c^i \nonumber\\
&& = {1\over 2} ~N(0)K(\omega)~ \epsilon_{ijk} ~
 \epsilon_{ijl} J^i J^j \otimes S_c^l~ S^k. 
\eea
The third order diagram gives 
\bea
&& [N(0)]^2 K(\omega) ~ S^kS^jS^i ~ 
  \mbox{Tr} [J^k\otimes S_c^k J^i\otimes S_c^i] ~J^j\otimes S_c^j 
   \nonumber\\
&& = {1\over 2} ~ [N(0)]^2 K(\omega) ~ \mbox{Tr} [J^iJ^i] ~
     J^j\otimes S_c^{j} ~S^iS^jS^i \nonumber\\
&& = {1\over 2} ~ [N(0)]^2 K(\omega) ~ \mbox{Tr} [J^iJ^i] ~
     J^j\otimes S_c^{j} ~[~i\epsilon_{ijk} S^kS^i + S^jS^iS^i ~]. 
\eea
Then after the self energy correction, third order scaling equations are
\bea
{\partial g^{i} \over \partial x}
 &=& {1\over 2} \left[ g^jg^k + g^kg^j \right]
    - {1\over 4} ~g^i ~ \mbox{Tr} \left[ g^jg^j + g^kg^k \right].
\eea
Here $(i,j,k)$ are cyclic.

\table

{\bf Table I}. Crystal electric field energy eigenstates for $J=5/2$
multiplet in the cubic symmetry.
\begin{center}
\begin{tabular}{|c||c|}
\hline
 $J=5/2$ Multiplet & States  \\[0.05in] \hline
$ |\Gamma_7^{(5/2)}; \up / \down > $
 & $- \sqrt{1\over 6} ~ |\pm 5/2> + \sqrt{5\over 6} ~|\mp 3/2>$  \\[0.05in]
\hline
$ |\Gamma_8^{(5/2)}; +, \up / \down > $
 & $ |\pm 1/2>$ \\[0.05in] \hline
$ |\Gamma_8^{(5/2)}; -, \up / \down > $
 & $\sqrt{5\over 6} ~|\pm 5/2> + \sqrt{1\over 6} ~|\mp 3/2>$ \\[0.05in] \hline
\end{tabular}
\end{center}

\begin{table}
\caption{Crystal electric field energy eigenstates for $J=7/2$
multiplet in the cubic symmetry.}\label{J3.5}
\begin{center}
\begin{tabular}{|c||c|}
\hline
$J=7/2$ Multiplet & States \\[0.05in] \hline
$ |\Gamma_6^{(7/2)}; \up / \down > $
 & $\pm \sqrt{5\over 12} ~|\mp 7/2> \pm \sqrt{7\over 12}~|\pm 1/2>$ \\[0.05in]
  \hline
$ |\Gamma_7^{(7/2)}; \up / \down > $
 & $ \pm {\sqrt{3} \over 2} ~|\pm 5/2> \mp {1\over 2}~ |\mp 3/2>$  \\[0.05in]
   \hline
$ |\Gamma_8^{(7/2)}; +, \up / \down  > $
 & $\pm \sqrt{7\over 12} ~|\mp 7/2> \mp \sqrt{5\over 12}~|\pm 1/2>$ \\[0.05in]
   \hline
$ |\Gamma_8^{(7/2)}; -, \up / \down > $
 & $\pm {1\over 2} ~|\pm 5/2> \pm {\sqrt{3} \over 2} ~|\mp 3/2>$ \\[0.05in]
\hline
\end{tabular}
\end{center}
\end{table}

\begin{table}
\caption{Decomposition of $J$-multiplets in cubic symmetry. All possible
$J$-multiplets are listed for the $f^2$ configuration.}\label{f2J}
\begin{center}
\begin{tabular}{|c||c|} \hline
 $J$-multiplets $D_J$ & Decomposition \\ \hline
 $D_{0}$ & $\Gamma_1$ \\
 $D_{1}$ & $\Gamma_4$ \\
 $D_{2}$ & $\Gamma_3 \oplus \Gamma_5$ \\
 $D_{3}$ & $\Gamma_2 \oplus \Gamma_4 \oplus \Gamma_5$ \\
 $D_{4}$ & $\Gamma_1 \oplus \Gamma_3 \oplus \Gamma_4 \oplus \Gamma_5$ \\
 $D_{5}$ & $\Gamma_3 \oplus 2 \Gamma_4 \oplus \Gamma_5$ \\
 $D_{6}$ & $\Gamma_1 \oplus \Gamma_2 \oplus \Gamma_3 \oplus \Gamma_4
    \oplus 2\Gamma_5$ \\
\hline
\end{tabular}
\end{center}
\end{table}

\begin{table}
\caption{Crystal electric field energy eigenstates for $J=4$
multiplet in the cubic symmetry.}\label{J4}
\begin{center}
\begin{tabular}{|c||c|}
\hline
$J=4$ Multiplet & States \\ \hline
$ | \Gamma_1^{(4)} > $
 & $\sqrt{5\over 24} ~|4> +\sqrt{7\over 12} ~|0> + \sqrt{5\over 24}
~|-4>$
   \\ \hline
$ | \Gamma_3^{(4)}; + > $
 & $ {1\over \sqrt{2} } ~(~|2> + |-2> ~)$  \\ \hline
$ | \Gamma_3^{(4)}; - > $
 & $ -\sqrt{7\over 24}~ |4> +\sqrt{5\over 12}~ |0> - \sqrt{7\over 24}~
|-4>$
  \\ \hline
$ | \Gamma_4^{(4)}; 0 > $
 & $ {1\over \sqrt{2} } ~(~|4> - |-4> ~)$  \\ \hline
$ | \Gamma_4^{(4)}; \pm 1 > $
 & $ {1\over \sqrt{8}} ~|\mp 3> + \sqrt{7\over 8} ~|\pm 1>$  \\ \hline
$ | \Gamma_5^{(4)}; 0 > $
 & ${1\over \sqrt{2} } ~(~|2> - |-2> ~)$ \\ \hline
$ | \Gamma_5^{(4)}; \pm 1 > $
 & $\sqrt{7\over 8} ~|\mp 3> - {1\over \sqrt{8}}~|\pm 1>$ \\
\hline
\end{tabular}
\end{center}
\end{table}

\begin{table}
\caption{Irreducible tensor operators for the conduction electrons.
As noted in the text, the $\Gamma_4$ irrep is the spin operator. Hence
there are all in all 6 possible conduction spin operators.
The mixing between $\Gamma_6$ and $\Gamma_7$ ($\Gamma_6 \otimes \Gamma_7$
$=\Gamma_2\oplus\Gamma_5$) does not have a spin degree of freedom.}
\label{tensor}
\begin{center}
\begin{tabular}{|c||c|} \hline
 $|~\Gamma^{\rm cb}~> \otimes <~\Gamma^{\rm cb}~|$ &
   Irreducible tensor operators \\ \hline
 $\Gamma_6 \otimes \Gamma_6$ & $\Gamma_1 \oplus \Gamma_4$ \\
 $\Gamma_7 \otimes \Gamma_7$ & $\Gamma_1 \oplus \Gamma_4$ \\
 $\Gamma_8 \otimes \Gamma_8$ & $\Gamma_1 \oplus \Gamma_2 \oplus
   \Gamma_3 \oplus 2\Gamma_4 \oplus 2\Gamma_5$ \\ \hline
 $\Gamma_6 \otimes \Gamma_8$ & $\Gamma_3 \oplus \Gamma_4 
   \oplus \Gamma_5$\\
 $\Gamma_7 \otimes \Gamma_8$ & $\Gamma_3 \oplus \Gamma_4 
   \oplus \Gamma_5$\\
\hline
\end{tabular}
\end{center}
\end{table}

\begin{table}
\caption{Selection rules for the Anderson hybridization between
$f^1\Gamma_7$
and $f^2$ configurations. For example, we can read off from this table
that only the $\Gamma_8$ conduction electrons can mix with $f^1\Gamma_7$
states to reach $f^2\Gamma_3$ states.}\label{hyb}
\begin{center}
\begin{tabular}{|c||c|} \hline
 $\Gamma^{\rm cb} \otimes \Gamma_{7}(f^1)$ & $f^2$ CEF states \\
  \hline
 $\Gamma_6 \otimes \Gamma_7$ & $\Gamma_2 \oplus \Gamma_5$ \\
 $\Gamma_7 \otimes \Gamma_7$ & $\Gamma_1 \oplus \Gamma_4$ \\
 $\Gamma_8 \otimes \Gamma_7$ & $\Gamma_3 \oplus \Gamma_4 \oplus \Gamma_5$
\\
  \hline
\end{tabular}
\end{center}
\end{table}

\begin{table}
\caption{We tabulate $<j_c\Gamma_7; {5\over 2} \Gamma_7| J_2\Gamma_1>$.
An asterisk denotes the violation of the triangular inequality 
for the addition of
two angular momenta. The $\Gamma_1$ irrep exists only in the $J=0,4,6$
multiplets in the $f^2$ configuration. }
\label{gamma1}
\begin{center}
\begin{tabular}{|c|c|c||c|} \hline\hline
 $j_c$  &  $J_1$  &  $J_2$  &  $<j_c\Gamma_7;{5\over 2}\Gamma_7|J_2\Gamma_1>$
  \\ \hline
 $5/2$  &  $5/2$  &  $0$    &  $\sqrt{1\over 6}$  \\
        &         &  $4$    &  $\sqrt{1\over 3}$  \\
        &         &  $6$    &  *  \\ \hline
 $7/2$  &         &  $0$    &  *  \\
        &         &  $4$    &  $\sqrt{5\over 22}$  \\
        &         &  $6$    &  $\sqrt{3\over 11}$  \\
\hline\hline
\end{tabular}
\end{center}
\end{table}

\begin{table}
\caption{We tabulate $<j_c\Gamma_6; {5\over 2} \Gamma_7| J_2\Gamma_2>$.
The $\Gamma_6$ irrep is absent in the $j_c=5/2$ partial waves.
The $\Gamma_2$ irrep exists only in the $J=3,6$
multiplets in the $f^2$ configuration.}
\label{gamma2}
\begin{center}
\begin{tabular}{|c|c|c||c|} \hline\hline
 $j_c$  &  $J_1$  &  $J_2$  &  $<j_c\Gamma_6;{5\over 2}\Gamma_7|J_2\Gamma_2>$
  \\ \hline
 $7/2$  &  $5/2$  &  $3$    &  $\sqrt{7/18}$  \\
        &         &  $6$    &  ${1\over 3}$  \\
\hline\hline
\end{tabular}
\end{center}
\end{table}

\begin{table}
\caption{Table for $<j_c \Gamma_8; {5\over 2} \Gamma_7 |J_2 \Gamma_3>$.
Note that $\sum_{J_2} |<j_c \Gamma_8; {5\over 2} \Gamma_7 |J_2 \Gamma_3>|^2
=0.5$, which is none other than a correct normalization.
The $\Gamma_3$ irrep exists only in the $J=2,4,5,6$ multiplets in the 
$f^2$ configuration.}
\label{gamma3}
\begin{center}
\begin{tabular}{|c|c|c||c|} \hline\hline
 $j_c$  &  $J_1$  &  $J_2$  &  $<j_c\Gamma_8;{5\over 2}\Gamma_7|J_2\Gamma_3>$
  \\ \hline
 $5/2$  &  $5/2$  &  $2$    &  $\sqrt{5\over 84}$  \\
        &         &  $4$    &  $-\sqrt{4\over 21}$  \\
        &         &  $5$    &  ${1\over 2}$  \\ \hline
 $7/2$  &         &  $2$    &  $\sqrt{25\over 126}$  \\
        &         &  $4$    &  $\sqrt{27\over 154}$  \\
        &         &  $5$    &  $\sqrt{1\over 18}$  \\
        &         &  $6$    &  $\sqrt{7\over 99}$  \\
\hline\hline
\end{tabular}
\end{center}
\end{table}

\begin{table}
\caption{Table for $< j_c \Gamma_7; {5\over 2} \Gamma_7 |J_2 \Gamma_4>$ and
$<j_c\Gamma_8;{5\over 2}\Gamma_7|J_2\Gamma_4>$ in Table \ref{gamma4}. 
Since the two $\Gamma_4$
irreps are present in the $J=5$ multiplets, the $J=5\Gamma_4$ CEF states 
are model parameter dependent. However, we note that the generic 
hybridization form is preserved\protect\cite{tskim}.}
\label{gamma4}
\begin{center}
\begin{tabular}{|c|c|c||c|c|} \hline\hline
 $j_c$  &  $J_1$  &  $J_2$  &  $<j_c\Gamma_7;{5\over 2}\Gamma_7|J_2\Gamma_4>$
 & $<j_c\Gamma_8;{5\over 2}\Gamma_7|J_2\Gamma_4>$
  \\ \hline
 $5/2$  &  $5/2$  &  $1$  & $-\sqrt{5\over 126}$  & ${2\over 3\sqrt{7}}$ \\
        &         &  $3$  & $-\sqrt{20\over 81}$ & ${1\over 18\sqrt{2}}$ \\
        &         &  $4$  & $0$                  & ${1\over \sqrt{8}}$ \\
        &         &  $5$  &   &                        \\ \hline
 $7/2$  &         &  $1$  & $-\sqrt{2\over 21}$ & $-\sqrt{1\over 28}$ \\
        &         &  $3$  & ${1\over \sqrt{24}}$ & $-{1\over 12}$  \\
        &         &  $4$  & ${7\over \sqrt{264}}$ & $-{1\over 4\sqrt{11}}$ \\
        &         &  $5$  &      &       \\
        &         &  $6$  & $\sqrt{7\over 66}$ & $-\sqrt{7\over 99}$  \\
\hline\hline
\end{tabular}
\end{center}
\end{table}

\begin{table}
\caption{Table for $<j_c \Gamma_{6,8}; {5\over 2} \Gamma_7 |J_2 \Gamma_5>$.
Since the two $\Gamma_5$ irreps are present in the $J=6$ multiplets, 
the CEF states are model parameter-dependent. However, we note that 
the generic hybridization form is preserved\protect\cite{tskim}.}
\label{gamma5}
\begin{center}
\begin{tabular}{|c|c|c||c|c|} \hline\hline
 $j_c$  &  $J_1$  &  $J_2$  &  $<j_c\Gamma_6;{5\over 2}\Gamma_7|J_2\Gamma_5>$
 & $<j_c\Gamma_8;{5\over 2}\Gamma_7|J_2\Gamma_5>$
  \\ \hline
 $5/2$  &  $5/2$  &  $2$  &  & $\sqrt{5\over 42}$ \\
        &         &  $3$  &  & $\sqrt{5\over 72}$ \\
        &         &  $4$  &  & $\sqrt{1\over 168}$ \\
        &         &  $5$  &  & $\sqrt{1\over 18}$  \\ \hline
 $7/2$  &         &  $2$  & $\sqrt{10\over 81}$ & $-{5\over 18\sqrt{7}}$ \\
        &         &  $3$  & $-{\sqrt{14}\over 36}$ & ${5\sqrt{5}\over 36}$  \\
        &         &  $4$  & $-\sqrt{5\over 264}$ &
                                          ${13\over 4\sqrt{3\cdot 77}}$ \\
        &         &  $5$  & ${\sqrt{70}\over 18}$   & ${1\over 9}$      \\
        &         &  $6$  &  &  \\
\hline\hline
\end{tabular}
\end{center}
\end{table}

\begin{table}
\caption{We list all the possible fixed points generated from the 
third order scaling equations of our model exchange interactions.
Most of them are unstable. Some of them might not exist in the full 
calculations.}\label{fixpt} 
\begin{center}
\begin{tabular}{|c|c|c|c|} \hline
 $g^{*}$ & $M^{i}$ & $\Lambda$ & Stability \\ 
  \hline
  $\infty$ & $\gamma^0$ & ${1\over 2} \cdot {3\over 2} ~ \gamma^0$ & Stable \\
  $\infty$ & $\kappa^0$ & ${1\over 2} \cdot {3\over 2} ~ \kappa^0$ & Stable \\
  $1$        & $\tau^0$   & ${1\over 2} \cdot {3\over 2} ~ \tau^0$ & Stable \\
  $2/3$      & $-{1\over 3} ~(~\gamma^0 + \tau^0~)
               \pm {2\sqrt{2} \over 3} ~ \gamma^{i} + {2\over 3} ~ \tau^{i}$ 
             & ${1\over 2} \cdot {3\over 2} ~(~\gamma^0 + \tau^0~)$ & Stable\\
  $2/3$      & $-{1\over 3} ~(~\kappa^0 + \tau^0~)
               \pm {2\sqrt{2} \over 3} ~ \kappa^{i} - {2\over 3} ~ \tau^{i}$ 
             & ${1\over 2} \cdot {3\over 2} ~(~\kappa^0 + \tau^0~)$ 
             & Stable\\[0.05in]
  \hline
  $0$        & $0$ & $0$ & Unstable \\
  $1/5$      & $-\tau^0 \pm 2~\tau^{i}$ & ${3\over 2}\cdot {5\over 2}~\tau^0$ 
             & Unstable \\
  $2/3$      & $\gamma^0 + \tau^0$ & ${1\over 2} \cdot {3\over 2}~
                (~\gamma^0 + \tau^0~)$ 
             & Unstable \\
  $2/3$      & $\kappa^0 + \tau^0$ & ${1\over 2} \cdot {3\over 2}~
               (~\kappa^0 + \tau^0~)$ 
             & Unstable \\
  $2/11$     & $\kappa^0 - \tau^0 \pm 2 ~ \tau^{i}$ 
             & ${1\over 2} \cdot {3\over 2}~\kappa^0
                + {3\over 2} \cdot {5\over 2} ~\tau^0$ & Unstable \\
  $2/11$     & $\gamma^0 - \tau^0 \pm 2 ~ \tau^{i}$ 
             & ${1\over 2} \cdot {3\over 2}~\gamma^0
                + {3\over 2} \cdot {5\over 2} ~\tau^0$ & Unstable \\
  $1/4$      & ${4\over 3}~\gamma^0 - {2\over 3}~ \tau^0
               \pm {2\sqrt{2} \over 3} ~ \gamma^{i} + {4\over 3} ~ \tau^{i}$ 
             & $1 \cdot 2 ~(~\gamma^0 + \tau^0~)$ & Unstable \\
  $1/4$      & ${4\over 3}~\kappa^0 - {2\over 3}~ \tau^0
               \pm {2\sqrt{2} \over 3} ~ \kappa^{i} - {4\over 3} ~ \tau^{i}$ 
             & $1 \cdot 2 ~(~\kappa^0 + \tau^0~)$ & Unstable \\
  $2/35$     & $- {5\over 3}~\gamma^0 + {7\over 3}~ \tau^0
               \pm {4\sqrt{5} \over 3} ~ \gamma^{i} + {4\over 3} ~ \tau^{i}$
             & ${5\over 2} \cdot {7\over 2} ~(~\gamma^0 + \tau^0~)$ 
             & Unstable \\
  $2/35$     & $- {5\over 3}~\kappa^0 + {7\over 3}~ \tau^0
               \pm {4\sqrt{5} \over 3} ~ \kappa^{i} - {4\over 3} ~ \tau^{i}$
             & ${5\over 2} \cdot {7\over 2} ~(~\kappa^0 + \tau^0~)$ 
             & Unstable \\
  $1$        & $\gamma^0 + \kappa^0$ 
             & ${1\over 2} \cdot {3\over 2}~(~\gamma^0 + \kappa^0~)$ 
             & Unstable \\
  $1/2$      & $\gamma^0 + \kappa^0 + \tau^0$ 
             & ${1\over 2} \cdot {3\over 2}~(~\gamma^0 + \kappa^0 + \tau^0~)$ 
             & Unstable \\
  $1/6$        & $\gamma^0 + \kappa^0 - \tau^0 \pm 2\tau^{i}$ 
             & ${1\over 2} \cdot {3\over 2}~(~\gamma^0 + \kappa^0~) 
                 + {3\over 2} \cdot {5\over 2} ~\tau^0$ & Unstable \\
  $1/2$      & $\kappa^0 - {1\over 3} ~(~\gamma^0 + \tau^0~)
               \pm {2\sqrt{2} \over 3} ~ \gamma^{i} + {2\over 3} ~ \tau^{i}$ 
             & ${1\over 2} \cdot {3\over 2} ~
                 (~\gamma^0 + \kappa^0 + \tau^0~)$ & Unstable \\
  $1/2$      & $\gamma^0-{1\over 3} ~(~\kappa^0 + \tau^0~)
               \pm {2\sqrt{2} \over 3} ~ \kappa^{i} - {2\over 3} ~ \tau^{i}$ 
             & ${1\over 2} \cdot {3\over 2} ~
                (~\gamma^0 + \kappa^0 + \tau^0~)$ & Unstable \\
  $2/9$      & $\kappa^0 + {4\over 3}~\gamma^0 - {2\over 3}~ \tau^0
               \pm {2\sqrt{2} \over 3} ~ \gamma^{i} + {4\over 3} ~ \tau^{i}$ 
             & ${1\over 2} \cdot {3\over 2}~\kappa^0 
                + 1 \cdot 2 ~(~\gamma^0 + \tau^0~)$ & Unstable \\
  $2/9$      & $\gamma^0 + {4\over 3}~\kappa^0 - {2\over 3}~ \tau^0
               \pm {2\sqrt{2} \over 3} ~ \kappa^{i} - {4\over 3} ~ \tau^{i}$ 
             & ${1\over 2}\cdot {3\over 2}~\gamma^0 
                 + 1 \cdot 2 ~(~\kappa^0 + \tau^0~)$ & Unstable \\
  $1/18$     & $\kappa^0 - {5\over 3}~\gamma^0 + {7\over 3}~ \tau^0
               \pm {4\sqrt{5} \over 3} ~ \gamma^{i} + {4\over 3} ~ \tau^{i}$
             & ${1\over 2}\cdot {3\over 2}~\kappa^0
                 + {5\over 2} \cdot {7\over 2} ~(~\gamma^0 + \tau^0~)$ 
             & Unstable \\
  $1/18$     & $\gamma^0 - {5\over 3}~\kappa^0 + {7\over 3}~ \tau^0
               \pm {4\sqrt{5} \over 3} ~ \kappa^{i} - {4\over 3} ~ \tau^{i}$
             & ${1\over 2}\cdot {3\over 2}~\gamma^0 
                + {5\over 2} \cdot {7\over 2} ~(~\kappa^0 + \tau^0~)$ 
             & Unstable \\
  $1/42$     & $3~\gamma^0 - {7\over 3}~\kappa^0 - {7\over 3}~\tau^0
                \pm 2\sqrt{3} ~ \gamma^{i}$ 
             & ${7\over 2} \cdot {9\over 2} ~
                (~\gamma^0 + \kappa^0 + \tau^0~)$ & Unstable \\
             & $\pm {2\sqrt{35} \over 3} ~ \kappa^{i} 
                + {4\over 3} ~ \tau^{i}$ & & \\
  $1/42$     & $-{7\over 3}~\gamma^0 + 3~\kappa^0 - {7\over 3}~\tau^0
               \pm {2\sqrt{35} \over 3} ~ \gamma^{i}$ 
             & ${7\over 2} \cdot {9\over 2} ~(~\gamma^0 + \kappa^0 + \tau^0~)$
             & Unstable \\
             & $\pm 2\sqrt{3} ~ \kappa^{i} - {4\over 3} ~ \tau^{i}$ 
             & & \\
\hline 
\end{tabular}
\end{center}
\end{table}

\begin{table}
\caption{Fractional parentage coefficients for 
$K(j_c; f^1{5\over 2}~|~f^2 LS=0:J)$. 
$*$ means the triangular sum is violated.
Note that for a given $(S,L,J)$, sum of the squared
fractional parentage coefficients over all possible $j_c = 5/2, 7/2$ is
always $6/7$.}\label{fracs}
\begin{center}
\begin{tabular}{|c||c|c|}
\hline
$(S,L,J)$  &     $j_c = 5/2$     &   $j_c = 7/2$       \\ \hline
$(0,0,0)$  &  $\sqrt{6\over 7}$  &    $*$             \\  \hline
$(0,2,2)$  &    ${6\over 7}$     & $\sqrt{6\over 49}$   \\ \hline
$(0,4,4)$  & $\sqrt{22\over 49}$ & $\sqrt{20\over 49}$  \\ \hline
$(0,6,6)$  &   $*$               & $\sqrt{6\over 7}$  \\
\hline
\end{tabular}
\end{center}
\end{table}

\begin{table}
\caption{Fractional parentage coefficients for
$K(j_c; f^1{5\over 2}~|~f^2 LS=1:J)$.
$*$ means the triangular sum is violated. 
Hund's rule ground multiplets
correspond to $(S=1,L=5,J=4)$.
Note that for a given $(S,L)$, sum of the squared
fractional parentage coefficients over all possible $j_c = 5/2, 7/2$
and over all $J=L, L\pm1$ is always $20/7$.}\label{fract}
\begin{center}
\begin{tabular}{|c||c|c|}
\hline
$(S,L,J)$  &     $j_c = 5/2$     &   $j_c = 7/2$       \\ \hline
$(1,1,0)$  &  $\sqrt{8\over 7}$  &    $*$              \\
$(1,1,1)$  &     $0$             &    $1$              \\
$(1,1,2)$  & $-\sqrt{8\over 49}$ & $-\sqrt{27\over 49}$ \\
\hline
$(1,3,2)$  & $\sqrt{54\over 49}$  & $-\sqrt{16\over 49}$ \\
$(1,3,3)$  &     $0$              & $1$   \\
$(1,3,4)$  & $-\sqrt{8\over 147}$ & $-\sqrt{55\over 147}$ \\
\hline
$(1,5,4)$  & $\sqrt{220\over 147}$ & $-\sqrt{32\over 147}$  \\
$(1,5,5)$  & $0$ & $1$  \\
$(1,5,6)$  & $*$ & $-\sqrt{1\over 7}$  \\
\hline
\end{tabular}
\end{center}
\end{table}

\begin{figure}
\protect\caption[Schematic display of all the possible exchange
interactions]
{{\bf Schematic display of all the possible exchange interactions.}
With the impurity pseudo-spin $S_I=1/2$ for $f^1J=5/2\Gamma_7$ doublet,
there are in total 6 exchange interactions for $\Gamma_6$,$\Gamma_7$,
and $\Gamma_8$ conduction electrons. In addition to the well-known
one-, and two-channel $S_c=1/2$ exchange interactions with
antiferromagnetic (AFM) coupling, there arise one-channel
$S_c=3/2$ exchange interactions with
antiferromagnetic coupling (not known in the literature), 
and one-channel $S_c=1/2$ exchange
interactions with ferromagnetic (FM) coupling. There are also
the mixing exchange interactions through which the conduction electrons
can change their CEF symmetry labels after interacting with the
impurity pseudo-spin.}
\protect\label{modelfig}
\end{figure}

\begin{figure}
\protect\caption[The scaling diagrams up to the third order]
{{\bf The scaling diagrams up to the third order.}
The solid lines denote the conduction electron propagator and the
dotted lines the impurity spin. The external solid lines are 
confined to the conduction electrons close to the Fermi level,
while the intermediate solid lines denote the conduction 
electrons virtually excited to the high energy states. 
At each vertex, the exchange
coupling matrices are determined by the specific exchange
interaction type. When the multiple conduction electron
partial waves are considered, the exchange coupling matrices
should be written as the direct product of the channel
matrix and the exchange interaction matrix.}
\protect\label{scalemat}
\end{figure}

\begin{figure}
\protect\caption[Schematic screening of $S_I=1/2$ by $S_c=3/2$]
{{\bf Schematic screening of $S_I=1/2$ by $S_c=3/2$.}
In the first shell screening, the energy is minimized by the compensation
of the impurity spin by the conduction electrons of $S_c=-3/2,-1/2$.
This picture does not violate the Pauli exclusion principle.
After the first shell screening, the effective impurity spin is
$S_I=3/2$.
In the second shell, $S_c=-3/2,-1/2$ conduction electrons cannot
come closer to the impurity site due to the Pauli exclusion principle.
Hence the effective exchange coupling becomes antiferromagnetic. 
In this case, the effective impurity spin is $S_I=1/2$. Thus
the effective impurity spins alternate between $S_I=1/2$ and 
$S_I=3/2$ with onion shell screening process. This alternating
picture will be verified in Chapter V using the numerical 
renormalization group technique.}
\protect\label{screen}
\end{figure}

\begin{figure}
\protect\caption[Flow diagram in the $\Gamma_8$ conduction electron
space]
{{\bf Flow diagram in the $\Gamma_8$ conduction electron space.}
There are three stable fixed points in the $\Gamma_8$ conduction electron
space. Only the two-channel fixed point remains stable when the full
exchange interactions are included. The other two fixed points
derived from $S_c=3/2$ conduction electrons become unstable
against the mixing exchange interactions between the
$\Gamma_{6,7}$ and the $\Gamma_8$ conduction electrons. See the text
for more details.}
\protect\label{g8flow}
\end{figure}

\end{document}